%% file: main_arxiv.tex
\documentclass[%
 aps, prl,
 amsmath,amssymb,
 superscriptaddress,
 preprint,%
]{revtex4-1}
\usepackage{graphicx}
\usepackage{dcolumn}
\usepackage{bm}
\usepackage{float}
\usepackage[utf8]{inputenc}
\usepackage[T1]{fontenc}
\usepackage{mathptmx}
\usepackage{amsmath}
\usepackage{textcomp} 
\usepackage{nicefrac}
\usepackage{xcolor}
\usepackage{siunitx} 
\definecolor{darkred}{rgb}{0.55,0,0}

\usepackage{ulem}
\begin{document}

\title[]{Verification of ultrafast spin transfer effects in FeNi alloys}
\newcommand{\exx}{\ensuremath{\epsilon_{xx}}}
\newcommand{\exy}{\ensuremath{\epsilon_{xy}}}
\newcommand{\reexy}{\ensuremath{\mathrm{Re}(\epsilon_{xy})}}
\newcommand{\imexy}{\ensuremath{\mathrm{Im}(\epsilon_{xy})}}
\newcommand{\vexy}{\ensuremath{\vec{\epsilon}_{xy}}}
\newcommand{\ef}{\ensuremath{E_\textsubscript{F}}}
\newcommand{\ptheta}{\ensuremath{\vec{p}_\theta}}
\newcommand{\permalloy}{Fe$_{19}$Ni$_{81}$}
\newcommand{\feni}{Fe$_{50}$Ni$_{50}$}

\author{Christina Möller}
\author{Henrike Probst}
\author{G. S. Matthijs Jansen} \email{gsmjansen@uni-goettingen.de} %
\author{Maren~Schumacher}
\author{Mariana~Brede}

\affiliation{I. Physikalisches Institut, Georg-August-Universit\"at G\"ottingen, Friedrich-Hund-Platz 1, 37077 G\"ottingen, Germany}

\author{John Kay Dewhurst}%
\affiliation{Max Planck Institute of Microstructure Physics, Weinberg 2, 06120 Halle, Germany}

\author{Marcel~Reutzel}
\author{Daniel Steil} %
\affiliation{I. Physikalisches Institut, Georg-August-Universit\"at G\"ottingen, Friedrich-Hund-Platz 1, 37077 G\"ottingen, Germany}

\author{Sangeeta Sharma} 
\affiliation{Max-Born-Institute for Non-linear Optics and Short Pulse Spectroscopy, Max-Born Strasse 2A, 12489 Berlin, Germany}

\author{Stefan Mathias} \email{smathias@uni-goettingen.de}%
\affiliation{I. Physikalisches Institut, Georg-August-Universit\"at G\"ottingen, Friedrich-Hund-Platz 1, 37077 G\"ottingen, Germany}

\begin{abstract}
The optical intersite spin transfer (OISTR) effect was recently verified in \feni{} using magneto-optical Kerr measurements in the extreme ultraviolet range. However, one of the main experimental signatures analyzed in this work, namely a magnetic moment increase at a specific energy in Ni, was subsequently found also in pure Ni, where no transfer from one element to another is possible. Hence, it is a much-discussed issue whether OISTR in FeNi alloys is real and whether it can be verified experimentally or not. Here, we present a comparative study of spin transfer in \feni{}, \permalloy{} and pure Ni. We conclusively show that an increase in the magneto-optical signal is indeed insufficient to verify OISTR. However, we also show how an extended data analysis overcomes this problem and allows to unambiguously identify spin transfer effects. Concomitantly, our work solves the long-standing riddle about the origin of delayed demagnetization behavior of Ni in FeNi alloys.



\end{abstract}

\maketitle


The ability to drive spin dynamics by ultrashort laser pulses offers a unique opportunity to bring the field of spintronics into the femtosecond regime, as demonstrated for example by the successful application of superdiffusive spin currents~\cite{Malinowski:2008co, rudolf_ultrafast_2012, battiato_superdiffusive_2010, melnikov_ultrafast_2011} in spintronic emitters~\cite{seifert_efficient_2016} and the possibility of direct magnetic phase switching by femtosecond laser pulses~\cite{Stanciu:2007fy, Lambert_Science_2014, schlauderer_temporal_2019}. Recently, a fascinating discovery was made in this context: given a suitable material, it is possible to drive spin transfer between sublattices directly by a strong optical field~\cite{hofherr_ultrafast_2020, tengdin_direct_2020, willems_optical_2020, siegrist_light-wave_2019, steil_efficiency_2020, Ryan.2023}.
This process, called optical intersite spin transfer (OISTR), even precedes the ultrafast demagnetization process and thus provides a path to even faster spintronic applications~\cite{el-ghazaly_progress_2020}.

OISTR was first proposed theoretically by the Sharma group \cite{Dewhurst2018} and has meanwhile been experimentally verified in a number of experiments. In the OISTR process, an ultrafast optical excitation drives spin-preserving electronic transitions from below the Fermi-level to above the Fermi-level. Due to the exchange-split bands in ferromagnets, different numbers of spin-up and spin-down electrons are excited by the laser pulse, which leads to an ultrafast spectral redistribution of the electron density as well as to an ultrafast spectral redistribution of the spins and thus to ultrafast dynamics in the spin-polarization. If the initial states for such an excitation are predominantly found in one elementary subsystem, e.g. of an alloy, and the final states are predominantly found in another elementary subsystem, an intersite spin-transfer occurs, which offers the potential for an advanced and extremely fast control of magnetic behaviour.

\begin{figure}[htb!] 
\centering
    \includegraphics[width=0.7\columnwidth]{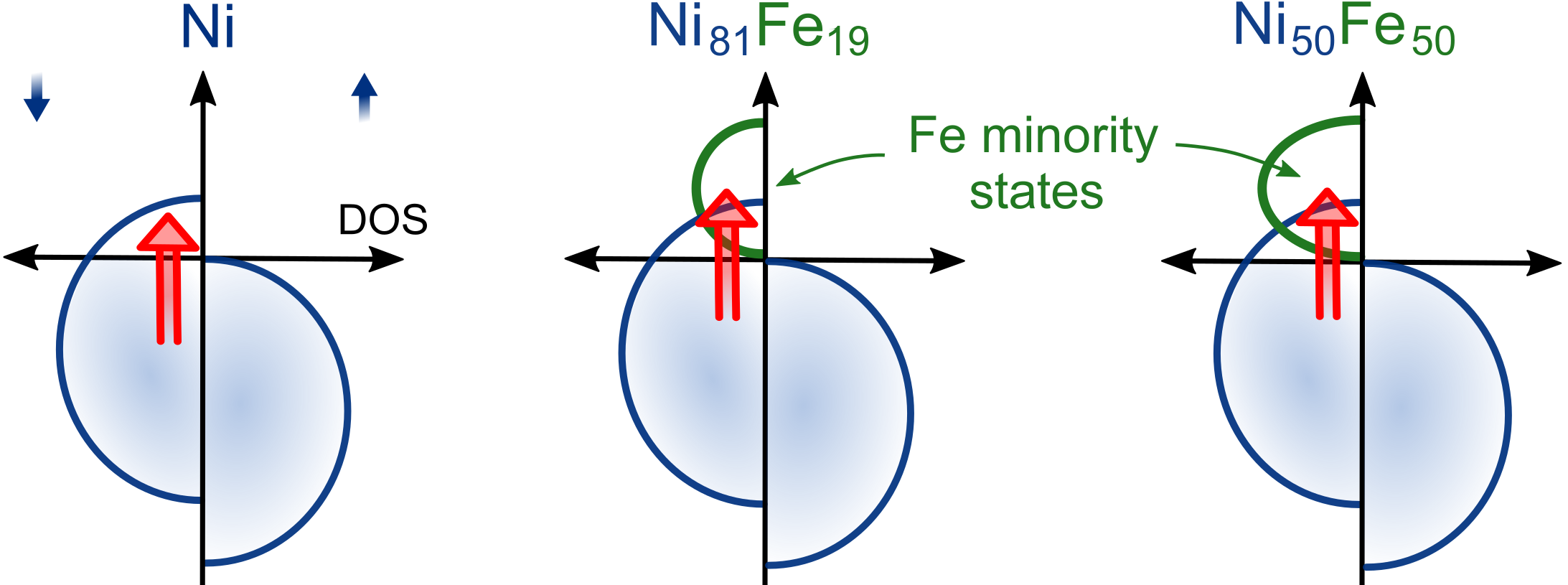}
    \caption{Simplified illustration of the 3d states in Ni, \permalloy{} and \feni{}. The addition of Fe in the alloys leads to additional unoccupied states in the minority channel that can be populated through OISTR. 
    TDDFT calculations predict a stronger OISTR effect in \feni{} compared to \permalloy{} due to the larger amount of available Fe states above the Fermi level \cite{Minar_2014}. On the other hand, in elemental Ni, the OISTR process is not possible.}
    \label{fig:simple_dos}
\end{figure}

One of the first experiments to verify the OISTR effect was an experiment on an \feni{} alloy using the ultrafast transverse magneto-optical effect (T-MOKE) in the extreme ultraviolet (EUV) region \cite{hofherr_ultrafast_2020}. The T-MOKE experiment showed a spectrally-dependent increase in magnetic asymmetry in Ni and a concomitant decrease in Fe, and the corresponding transient dynamics of majority and minority spin occupation in the \feni{} alloy were calculated by TDDFT. However, a similar ultrafast increase in Ni has recently also been observed in pure Ni material \cite{hennes_time-resolved_2021, probst_unraveling_2023}, where spin transfer to another subsystem is not possible. On the one hand, such a signature is not unexpected, since ultrafast excitation also drives occupation changes in pure materials. On the other hand, the question arises to what extent the observed increase in the magnetic asymmetry is a valid signature to verify spin transfer from Ni to the Fe in \feni{}. The situation is further complicated by the fact that it was found that identical magnetization dynamics in EUV T-MOKE can appear drastically different depending on the precise experimental geometry~\cite{probst_unraveling_2023}. 
The critical question is therefore whether OISTR can experimentally be verified in \feni{}.

\begin{figure*}[tbh!] 
\centering
    \includegraphics[width=1\textwidth]{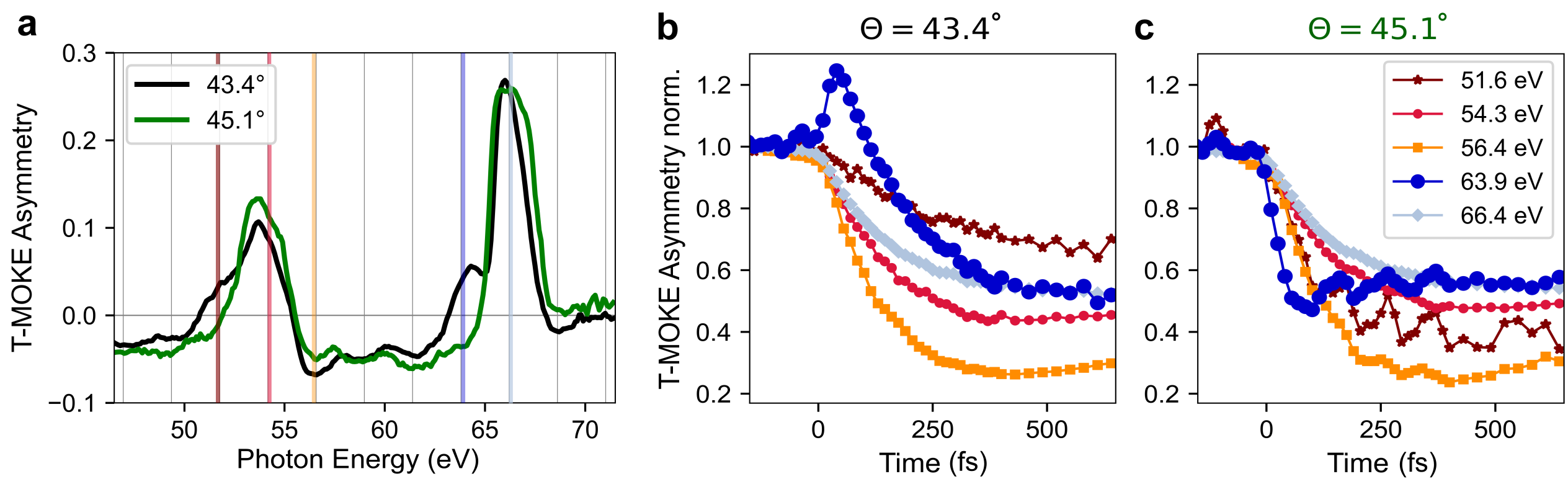}
    \caption{Apparently contradictory EUV T-MOKE measurements of OISTR in \permalloy{}, recorded at different EUV incidence angles and otherwise identical conditions (the data of \feni{} and Ni can be found in the Supplementary Figure~\ref{fig:all_HH_all}). (a) Static T-MOKE asymmetry for two incidence angles. Note that the magnetic asymmetry changes sign around 52~eV and 64~eV for the two different incidence angles. (b, c) Time-resolved magnetic asymmetry traces for 43.4\textdegree{} and 45.1\textdegree{} incidence angle, respectively. The analyzed photon energies around the M absorption edges of Fe and Ni are marked as colored bars in (a).  At 43.4\textdegree{}, the typical transient increase at 63.9~eV that was previously associated with OISTR is seen \cite{hofherr_ultrafast_2020}, while no such increase is visible at 45.1\textdegree{}.
    }
    \label{fig:tmoke_contradiction}
\end{figure*}

In this work, we present new data on this topic in a comparative experimental and theoretical study of laser-driven spin transfer processes in \feni{}, \permalloy{} (permalloy), and pure Ni (see Fig.~\ref{fig:simple_dos}). We revisit the previously observed signatures of the OISTR effect and, following the approach that we developed in Ref.~\onlinecite{probst_unraveling_2023}, perform a full time-resolved reconstruction of the dielectric tensor. This allows us to directly compare the same quantity obtained from experiment and theory: namely, the transient dynamics in the dielectric tensor. We find that in all three materials signatures of the ultrafast laser-driven spectral redistribution of spins can be clearly identified. 
Furthermore, we find that the amplitude of these dynamics at the Ni site increases with increasing Fe content in the alloy, supporting the OISTR description. 
In the case of \feni{}, we experimentally verify the transfer of minority spins from the Ni subsystem to the Fe subsystem, i.e. the OISTR effect.
Finally, we are also able to explain the origin of the much discussed delayed demagnetization behavior of Fe and Ni in these alloys \cite{Mathias2012, Guenther2014, yao_distinct_2020, Jana2017, moller_ultrafast_2021, Jana2022}.

\section{Results} 
We begin with a measurement of \permalloy{} that very clearly illustrates that an increase in T-MOKE asymmetry in a specific spectral region may not be sufficient to verify spin transfer effects without further analysis. Fig.~\ref{fig:tmoke_contradiction}a shows energy-resolved magnetic asymmetries measured with EUV T-MOKE of the \permalloy{} sample for two different incidence angles ($\theta$) of the EUV light. While the asymmetries look quite similar, the zero crossings are slightly shifted so that there are spectral regions where the asymmetry is negative for one angle of incidence and positive for the other, most notably at 52 and 64~eV. Most curiously, we find apparently contradictory ultrafast dynamics of the T-MOKE asymmetry in the time-resolved experiment for these spectral regions: Fig.~\ref{fig:tmoke_contradiction}b shows the time-resolved change of the asymmetry as a function of EUV energy at $\theta =$~43.4\textdegree{}. Here, the signal at 63.9~eV (dark blue) shows a time-resolved increase in the T-MOKE asymmetry, which is exactly the signature used to verify the OISTR effect in our previous work. However, in Fig.~\ref{fig:tmoke_contradiction}c, for a slightly different angle of incidence ($\theta =$~45.1\textdegree{}), the same spectral region shows the exact opposite behavior, namely an ultrafast decrease. Note that we observe this behavior for all three samples Ni, \permalloy{}, and \feni{} (cf. Supplementary Figure~\ref{fig:all_HH_all}). Clearly, no conclusion should be drawn from these data without further knowledge of the OISTR effect and its signatures in the T-MOKE signal.

To overcome this problem, we apply an extended analysis, based on the experimental determination of the transient dynamics of the dielectric tensor that we developed in Ref.~\cite{probst_unraveling_2023}. This analysis shows that the peculiar behavior of the asymmetry increase and decrease for two different incidence angles is due to a transient rotation of the off-diagonal element of the dielectric tensor (\exy{}) in the complex plane. If this rotation of \exy{} happens at photon energies near the zero-crossings of the T-MOKE asymmetry, it can lead to opposite dynamics in the magnetic asymmetry due to the projection of \exy{} onto an angle-dependent probe vector (see Ref.~\cite{probst_unraveling_2023} for the description of the analysis and the influence of the rotation of \exy{} on the T-MOKE asymmetry spectra; see SI for the additional data needed to perform this analysis). 
Instead, extracting \exy{} from the angle-dependent T-MOKE asymmetry data has several advantages: (i) the real part of \exy{} is related to the spin-polarization of the unoccupied states and can also be measured by other techniques, such as X-ray magnetic circular dichroism (XMCD); (ii) the quantity \exy{} is now independent of the measurement technique or geometry used; (iii) the transient changes in \exy{} allow a direct comparison with TDDFT calculations.

\begin{figure*}[btp!] 
\centering
    \includegraphics[width=1\textwidth]{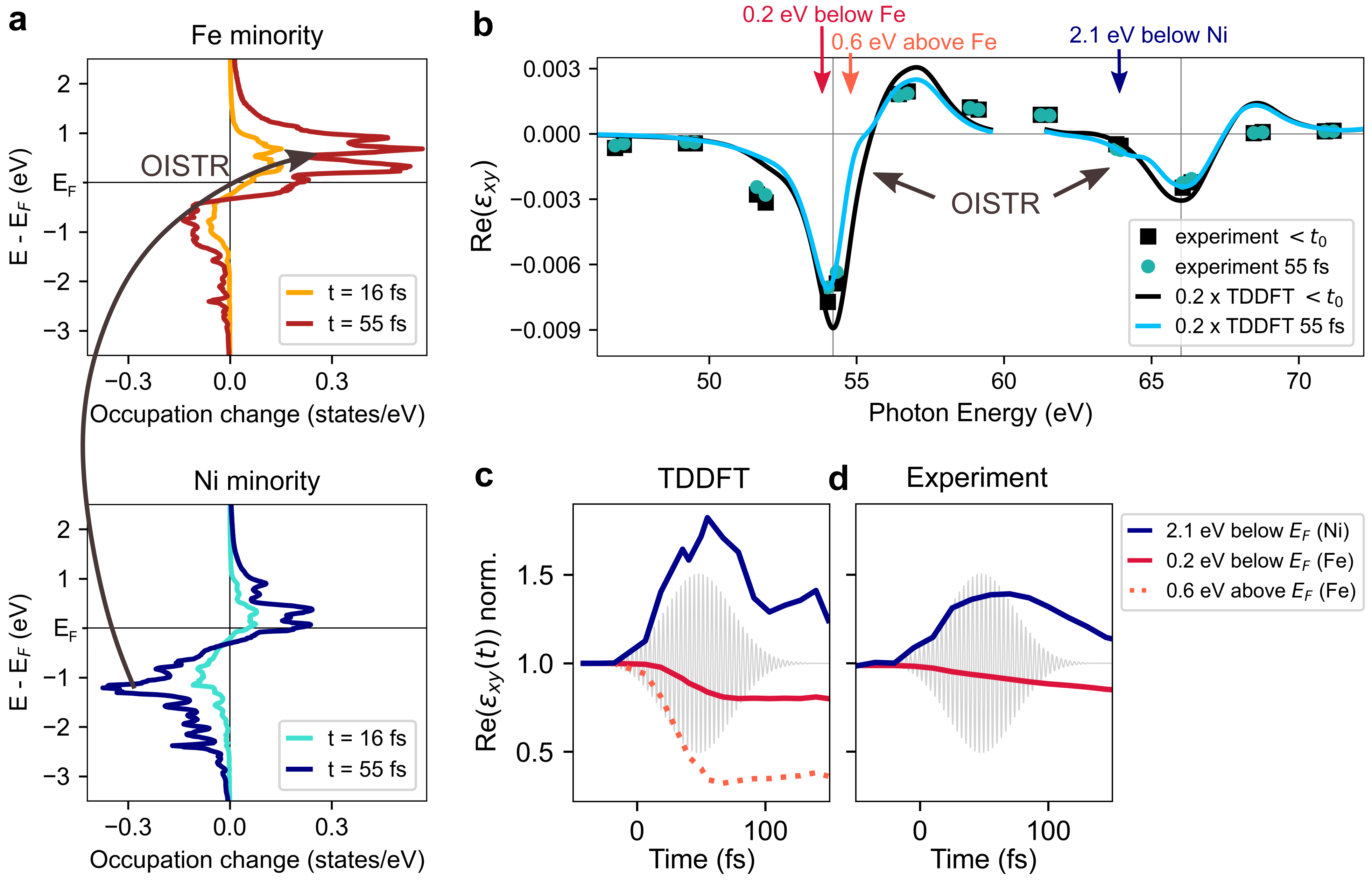}
    \caption{TDDFT calculations of OISTR in \feni{}, compared to experiment. a) The transient occupation change of minority spins at the Ni and Fe sites in \feni{}. The OISTR process excites minority spins in Ni from below the Fermi edge to Fe minority states above the Fermi edge. b) Comparison of the static off-diagonal dielectric tensor element to the transient state at 40~fs, as extracted from measured EUV T-MOKE data (points) and TDDFT calculations (lines). For the red, orange, and blue arrows, 54.2~eV and 66~eV were used for the M edges of Fe and Ni, respectively, which allows to reference the energy scale to the Fermi level. In (c) and (d), the transient evolution of the real part of \exy{} is shown at specific energies referenced to the Fermi level from TDDFT and experiment, respectively. For quantitative comparison, \exy{}(t) was normalized to its value before the pump pulse arrives.
   }
    \label{fig:exy_re}
\end{figure*}

Fig.~\ref{fig:exy_re}a shows TDDFT calculations~\cite{Dewhurst.2016, krieger2015laser} of the transient dynamics of the spin-resolved occupation in Fe and Ni in the \feni{} sample. A calculation of \permalloy{} is computationally very costly due to the large required supercell size and was therefore not performed for the present study. The calculations for \feni{} are similar to the results presented in Hofherr \textit{et al.} \cite{hofherr_ultrafast_2020}, but adapted for the pump pulse energy (1.2~eV) and pulse duration (47~fs) of the present experiment \cite{Bierbrauer:2017db}. While the Fe and Ni electrons share a common (metallic) band structure in the alloy, the character of the different states can be projected onto bands originating dominantly from the Ni or Fe subsystem. The OISTR effect then manifests itself in transitions from minority states below the Fermi level with predominantly Ni character to minority states above the Fermi level with predominantly Fe character. Of course, other transitions within the subsystems also contribute to the optical response, but they do not transfer spin from one subsystem to the other subsystem and are therefore not discussed further here. Transitions between the subsystems within the majority channel are also possible, but these are only minor changes compared to the strong changes in the minority channel (cf. Supplementary Figure~\ref{fig:TDDFT_DOS}).
As can be seen from these calculations, there is a spectral region below the Fermi level that shows a depletion of minority spins for states with a predominant Ni character, leading to an increase in the energy-resolved magnetic moment at the Ni site. Conversely, there are spectral regions above the Fermi level of the Fe states where the increase in minority spins leads to a rapid quenching of the energy-resolved magnetic moment. Hofherr \textit{et al.} \cite{hofherr_ultrafast_2020} found two similar features in the time-resolved asymmetry and interpreted them as indicators of the OISTR effect. 

However, for the reasons given above, we now go one step further and calculate from TDDFT the signature of OISTR on the transient dynamics of the dielectric tensor~\cite{Dewhurst2020PRL, dewhurstsharma}. Fig.~\ref{fig:exy_re}b shows the real part of \exy{} from theory (lines) and experiment (points) for two exemplary pump-probe delays (before time zero and at 55~fs). The first important observation from the calculated time- and spectrally-resolved \reexy{} data is that the spectrally very distinct dynamics in the spin-resolved DOS (Fig.~\ref{fig:exy_re}a) are strongly broadened. This is due to the intrinsic linewidth of the 3p core levels and their partial overlap. Nevertheless, the second important observation is that it is still possible to pinpoint the OISTR effect from the theory data: For photon energies between 63.0~eV and 64.4~eV, there is a relative increase in \reexy{} caused by the pump-induced ultrafast loss of minority spins in Ni. Conversely, in the 55~eV energy region, Fe shows a rapid decrease in \reexy{} on the timescale of the pump pulse due to the addition of transferred minority spins originating from Ni states.


Fig.~\ref{fig:exy_re}c summarizes the expected transient behavior of \reexy{} from TDDFT for the discussed energies, with the blue curve showing the increase in Ni and the orange dotted curve showing the corresponding rapid decrease in Fe, now approximately referenced to the Fermi levels of Ni and Fe, respectively, by subtracting the respective M-edge energies from the photon energies (cf. Fig.~\ref{fig:exy_re}b for the corresponding photon energies). In direct comparison, Fig.~\ref{fig:exy_re}d shows the experimentally extracted transient dynamics of \reexy{} for respective photon energies. First, we see that we are able to verify the important increase in \reexy{} in experiment, which is indicative of the loss of minority spins in Ni due to optical pumping (blue line). Second, however, we do not have sufficient EUV intensity in the important spectral region just above the Fe M-edge, which is expected to show the strong and rapid decrease due to the increase of minority spins above the Fermi level (orange dotted line in theory, Fig.~\ref{fig:exy_re}c, absent in experiment, Fig.~\ref{fig:exy_re}d). The next available EUV harmonic with sufficient intensity in our experimental data is at 0.2~eV below Fe M-edge, and we find good qualitative agreement with theory (red lines in Fig.~\ref{fig:exy_re}c,d). However, there is also a qualitative discrepancy with respect to timescales larger than 80~fs, where the theoretical magnetization remains constant while the experimental magnetization decreases. This can be understood by considering the limitations of the TDDFT calculations: TDDFT is well suited to describe the very early magnetization dynamics induced by the pump pulse and in particular the excitation. On longer timescales, however, the well-known demagnetization processes evolve, not all of which are included in TDDFT. In particular, Elliott-Yafet spin-flip electron-phonon scattering \cite{Koopmans:2010eu} is not included in TDDFT, which explains that TDDFT does not capture the full magnetization decrease for timescales >80~fs. In summary, we can clearly verify the loss of minority spins in the Ni subsystem from these experimental data, but we have no experimental evidence that these spins are transferred to the Fe subsystem.

\begin{figure}[htp!]
\centering
    \includegraphics[width=0.62\textwidth]{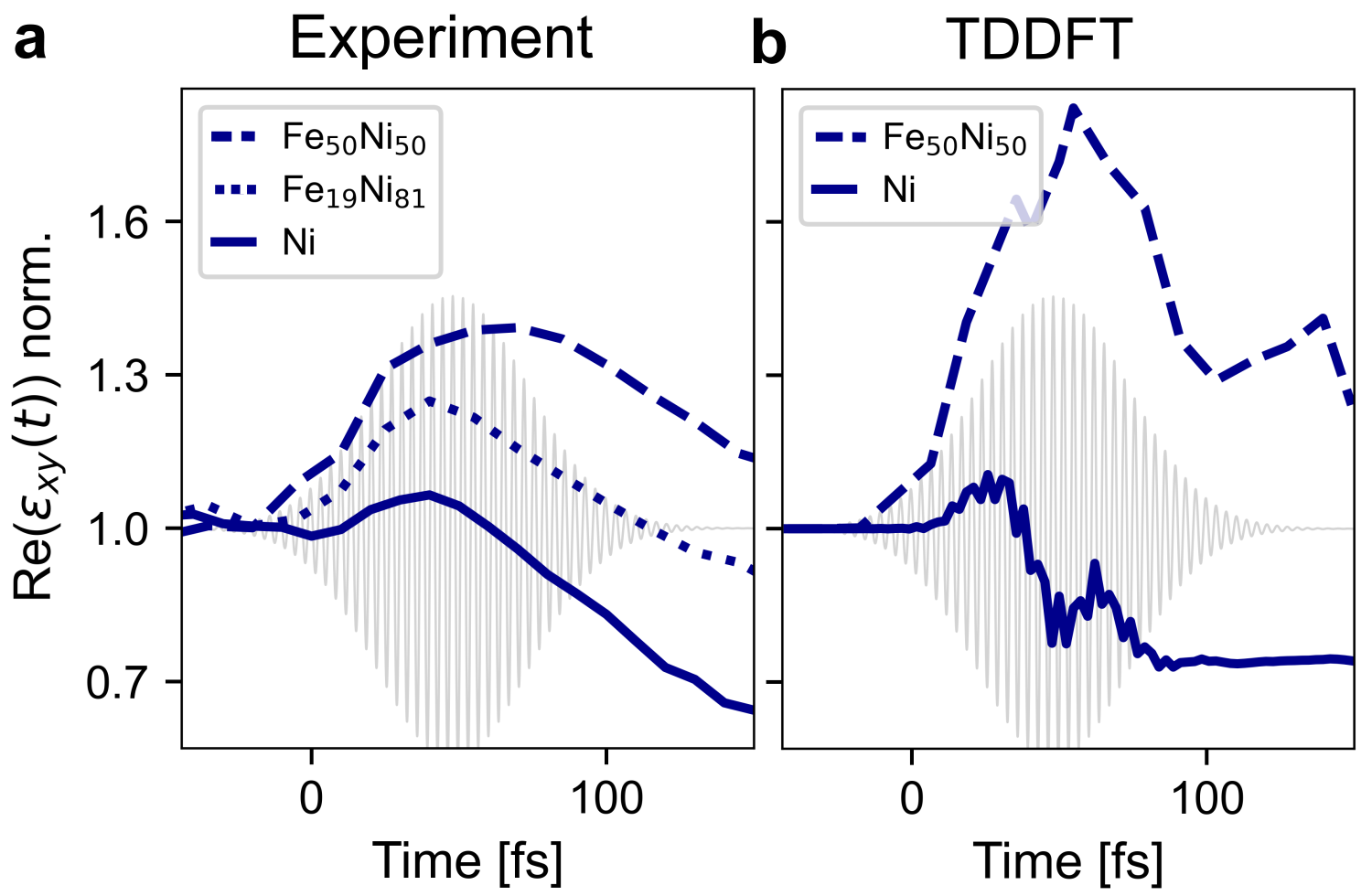}
    \caption{Comparison of the transient off-diagonal tensor element \reexy{} for 63.9~eV for \feni{}, \permalloy{} and Ni, (a) extracted from measured EUV T-MOKE data, and (b) calculated by TDDFT. The spin transfer between Ni and Fe is visible by an increase of \reexy{} at 63.9~eV, which probes Ni minority states below the Fermi level. In experiment, the spin transfer is found to be more efficient in \feni{} than in \permalloy{}. In Ni, minority electrons are excited from below to above the Fermi level, which results in a short and comparatively small increase of \reexy{}.
    }
    \label{fig:exy_re_all}
\end{figure}

For the present work, we now aim to compare the spectrally-resolved T-MOKE data of pure Ni \footnote{This data is reproduced from Ref.~\cite{probst_unraveling_2023}, where we describe the \exy{} analysis procedure in detail}, where only an inter-energy spin-transfer is involved, with data from \permalloy{} and \feni{}, where an intersite spin transfer between Ni and Fe becomes possible and is predicted by theory (for the \feni{} alloy). Fig.~\ref{fig:exy_re_all}a shows the experimentally analyzed transient change of \reexy{} for the spectral region where the OISTR-induced increase was expected and verified in the case of \feni{} (dashed). Compared to \feni{}, the increase in \permalloy{} (dotted) is much less pronounced and mostly seen as a delay in the case of pure Ni material (line). From these data, we can directly conclude that the OISTR-relevant transition is most efficiently excited in the \feni{}, less efficiently in \permalloy{}, and even less in Ni. In comparison with theory (Fig.~\ref{fig:exy_re_all}b), we again find good qualitative agreement for \feni{} as discussed above, but also for Ni. Note that the absorbed fluence and the quenching of the transient asymmetry are not the same for all three measurements (cf. SI).

\begin{figure}[htp!] 
\centering
    \includegraphics[width=0.65\textwidth]{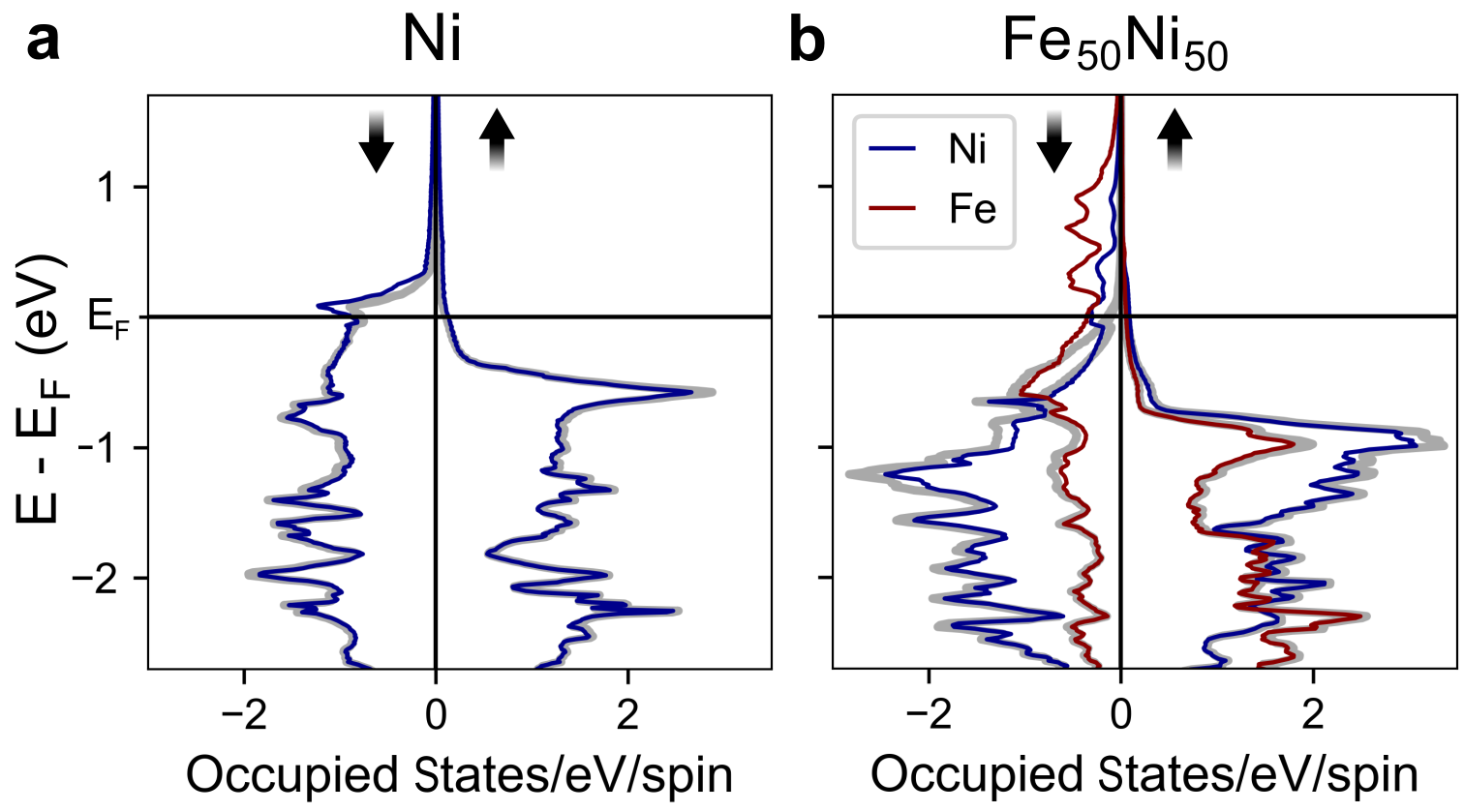}
    \caption{Occupied minority and majority states before (grey) and after the optical excitation (t=55~fs, blue and red) calculated with TDDFT for Ni (a) and \feni{} (b).}
    \label{fig:tddft_55fs}
\end{figure}

To shed light on the origin of the observed differences in excitation efficiencies in the minority channels, we take a closer look at the band-structure of these materials, and possible excitation pathways. Fig.~\ref{fig:tddft_55fs}a shows the spin-resolved density of states for pure Ni material. In Ni, more (spin-conserving optical) excitations are possible in the minority channel than in the majority channel. These transitions are captured in theory and experiment by a loss of minority spins below the Fermi level, and lead to the observed small increase for \reexy{} at $\approx$2.1~eV below the Ni edge. In the case of \feni{}, as shown in Fig.~\ref{fig:tddft_55fs}b, the situation is different. Here, the addition of Fe adds additional final states above the Fermi level in the minority channel. Thus, a first glance at the band structure suggests that additional transitions from the Ni minority states below the Fermi level to the Fe minority states above the Fermi level become possible. According to theory, this is indeed the case and leads to the strong relative increase of \reexy{} at 2.1~eV below the Ni edge and, in addition, to a rapid relative decrease of \reexy{} just above the Fe edge. While we cannot probe the rapid decrease with the given EUV light source in our experiment, we very well reproduce the modified efficiency of the OISTR transition and therefore conclude that the experiment also shows that in \feni{} OISTR is operative, i.e. minority spins from Ni are indeed pumped into minority states of Fe by the ultrafast laser excitation. Next, we see that the \permalloy{} alloy, according to the experiment (see Fig.~\ref{fig:exy_re_all}a), lies between the situation of Ni, where no intersite spin transfer is possible, and \feni{}, where OISTR is apparently strong. Using the same reasoning as above, one would expect some amount of Fe DOS above the Fermi level, but less than in the case of \feni{}. Indeed, such a series of spin-resolved DOS has already been calculated in Ref.~\cite{Minar_2014} and is in full agreement with the expectations and observations developed above. 

Finally, we would like to discuss our new results in relation to the results on permalloy obtained in the last decade with EUV T-MOKE (Mathias \textit{et al.} \cite{Mathias2012}, Günther \textit{et al.} \cite{Guenther2014} suppl., Jana \textit{et al.} \cite{Jana2017}). In these works, Fe and Ni showed a delayed demagnetization behavior with identical demagnetization time constants, which was recently verified in XMCD experiments at the L-edges of Fe and Ni \cite{Jana2022}. The identical demagnetization time constants for Fe and Ni aren't surprising: Fe and Ni are strongly exchange-coupled in these alloys. However, the important question has always been what causes the delay between the two subsystems. The interpretation of the OISTR process induced by the pump pulse now makes this clear. OISTR initiates a non-equilibrium between the Ni and Fe subsystems via the transfer of minority spins on the very short timescale of the pump excitation. Subsequently, demagnetization occurs via the well-known spin flip and exchange scattering processes, and both subsystems start to demagnetize at the same rate. At the time of the first experiment \cite{Mathias2012}, the small increase due to OISTR in permalloy was not resolved, but the subsequent delayed behaviour in the demagnetization of the Fe and Ni subsystems was clearly identified. 
However, for the reasons given above, we also revisit the delayed demagnetization of Ni and Fe and perform an energy-integrated analysis of \reexy{}. From this extended analysis, we still find a demagnetization delay for Ni in \permalloy{} of $12\pm3$~fs, while \feni{} actually shows both a transient increase in Ni and a relative delay of $95\pm 7$~fs (see SI for the transient dynamics of \reexy{}). 
We emphasize that it is really the stronger OISTR effect that enhances the delayed behavior, and not a modification of the exchange coupling that also influences the delayed behaviour, as was previously observed for permalloy alloyed with Cu \cite{Mathias2012}.

In conclusion, we have carried out a combined experimental-theoretical analysis of optically-driven spin-transfer in Ni, \permalloy{}, and \feni{}. We have paid special attention to the observed increase and delay of the magnetic asymmetry in T-MOKE, which was found in the alloys, but also in pure Ni material. Through an extended analysis (see Ref.~\cite{probst_unraveling_2023}), we are able to directly compare transient dynamics in the real part of the off-diagonal element of the dielectric tensor \reexy{}. 
Crucially, this procedure allows us to make a direct comparison with TDDFT theory calculations, which helps us to identify and explain all the observed signals. In summary, we verify OISTR in \feni{} and elucidate the origin of the previously found delays in these material systems. 

\input{methods}

\section{Data availability}
The data that support the findings of this study are available from the corresponding author upon reasonable request.

\section{References}
\bibliography{2022_FeNi}

\section{Acknowledgements}
We thank Mario Fix and Manfred Albrecht from the University of Augsburg for the fabrication of the FeNi samples. We also thank Thomas Brede from the Institut für Materialphysik Göttingen for the fabrication of the Ni sample.
This work was funded by the Deutsche Forschungsgemeinschaft (DFG, German Research Foundation) - project IDs 399572199  and 432680300/SFB 1456. G.S.M.J. acknowledges financial support by the Alexander von Humboldt Foundation. S.S. and J.K.D. would like to thank the DFG for funding through project-ID 328545488 TRR227 (project A04).

\section{Author contributions}
G.S.M.J., M.R., D.S., S.S. and S.M. conceived the research. C.M., H.P., M.S. and M.B. carried out the measurements. C.M., H.P. and G.S.M.J. performed the data analysis. J.K.D. and S.S. performed the TDDFT calculations. All authors discussed the results. G.S.M.J., D.S. and S.M. were responsible for the overall project direction. C.M., H.P., G.S.M.J. and S.M. wrote the manuscript with contributions from all co-authors. 

\section{Supplementary Information}
\renewcommand\thefigure{S\arabic{figure}} 
\setcounter{figure}{0}
\input{supple_text}

\end{document}

%% file: methods.tex
\section{Methods}


\section{EUV T-MOKE data for FeNi alloys and Ni}
\label{measurements}
We measured the transient T-MOKE asymmetry for multiple incidence angles in order to be able to extract the transient off-diagonal tensor element \reexy{}. The reflected 100~kHz EUV probe beam spans energies between 30-72~eV. We estimate the resolution of the spectrometer to be better than 0.2~eV, while the photon energy calibration is accurate to $<2\%$. Details on the angle-resolved measurement and analysis are given in Ref.~\cite{probst_analysis_2023}, while general information on the experimental setup can be found in Ref.~\cite{moller_ultrafast_2021}.

The experimental T-MOKE asymmetries and their dynamics for \feni{}, \permalloy{}  and Ni are shown in Supplementary Figure~1. Here, we pumped the samples with a $47 \pm 5$~fs pulses (Gauss FWHM) with a photon energy of 1.2~eV. The absorbed fluence is slightly different for each measurement: $0.8 \pm 0.2~\nicefrac{\text{mJ}}{\text{cm}^2}$ for Ni, $1.1 \pm 0.2~\nicefrac{\text{mJ}}{\text{cm}^2}$ for \feni{} and $0.8 \pm 0.2~\nicefrac{\text{mJ}}{\text{cm}^2}$ for \permalloy{}.

%% file: supple_text.tex
\section{EUV T-MOKE data for FeNi alloys and Ni}
\begin{figure*}[htp]
\centering
    \includegraphics[width=\textwidth]{{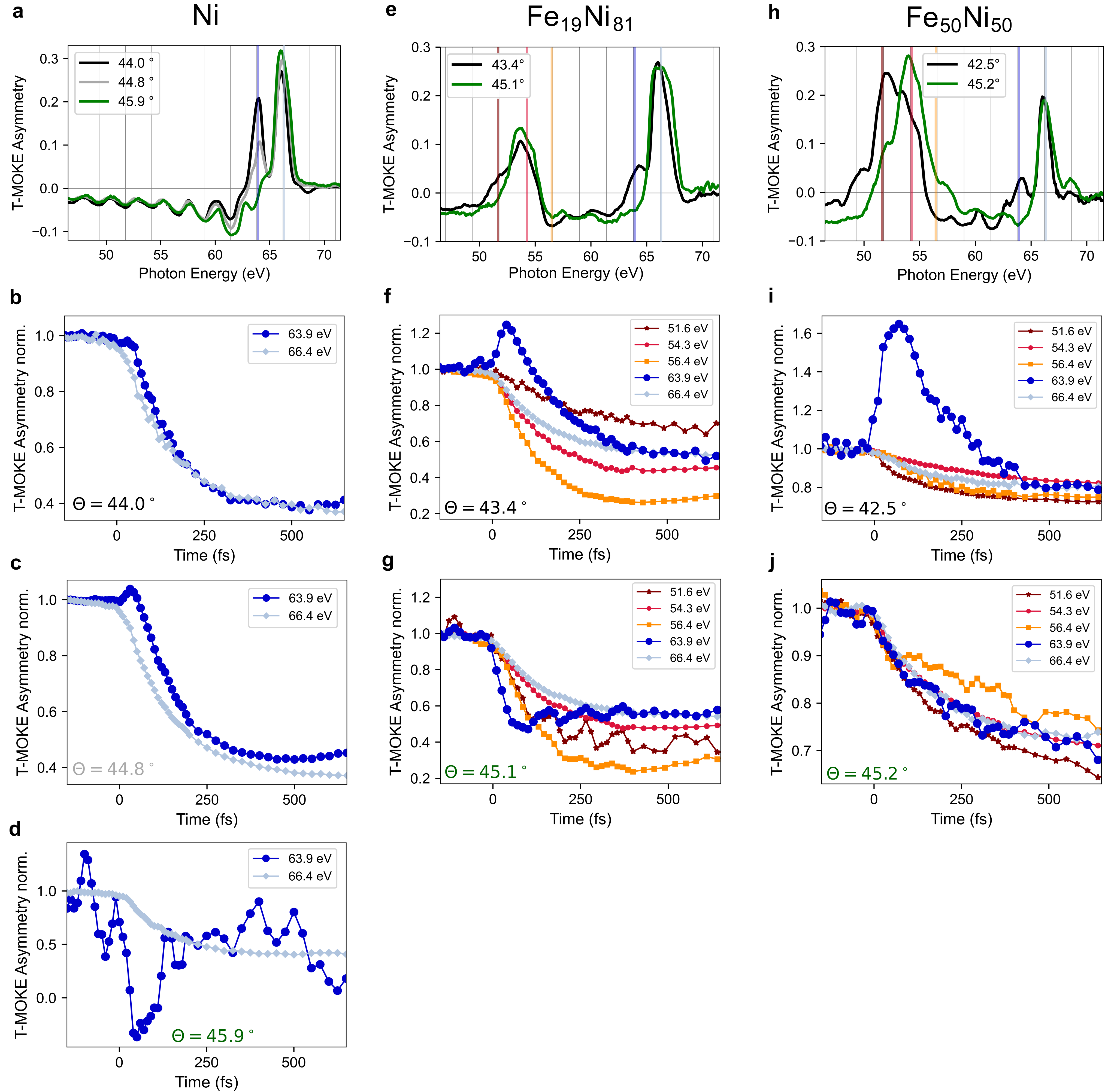}}
    \caption{All experimental results for Ni, \permalloy{} and \feni{}. We measured the transient T-MOKE asymmetry for three different incidence angles for Ni (a-d), and for two different incidence angles for the two FeNi alloys (e-j). The asymmetry around 63.9~eV photon energy shows dramatically different dynamics for different incidence angles.
   }
    \label{fig:all_HH_all}
\end{figure*}

Supplementary Figure~\ref{fig:all_HH_all} shows the complete data sets that were used to fit \reexy{}.

\newpage
\section{Delayed dynamics in FeNi alloys}
In order to extract the relative demagnetization delay, we evaluate the FeNi asymmetries for large energy regions around the M edge, as has been done in earlier T-MOKE studies \cite{Mathias2012, Guenther2014, Jana2017, moller_ultrafast_2021}. In contrast to the spectrally-resolved analysis in the main paper, we would call this an element-specific analysis of the data.

\begin{figure*}[tbh]
\centering
    \includegraphics[width=1\textwidth]{{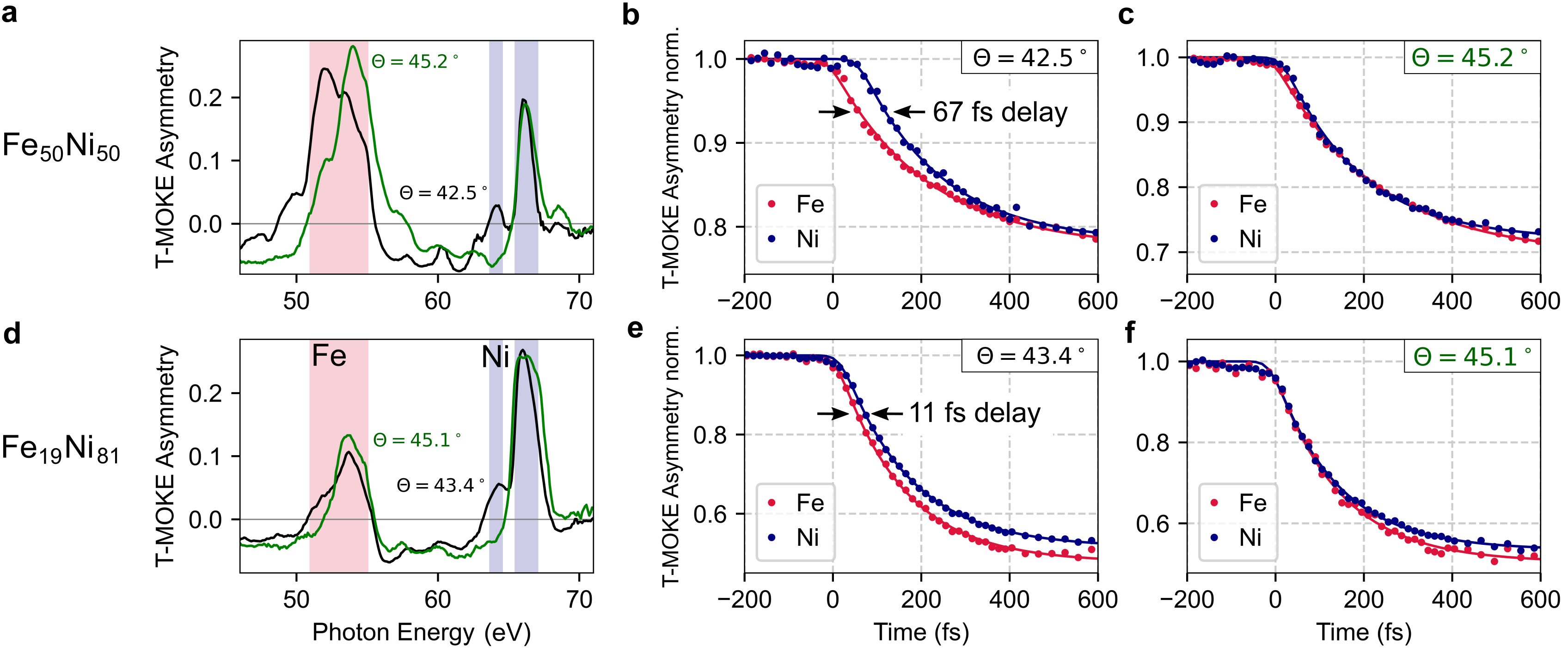}}
    \caption{Magnetic asymmetry for two incidence angles of a) \feni{} and d) \permalloy{}. b), c), e) and f) Transient evolution of the magnetic asymmetry integrated in the marked energy regions around each absorption edge: Fe: 51\,-\,55~eV, Ni: 63.7\,-\,64.5~eV and 65.5\,-\,67~eV. We observe the delayed behavior of Ni compared to Fe at the smaller incidence angle [b) and e)], while there is no such delay present for the transient asymmetry measured at the larger incidence angle. 
    }
    \label{fig:delay}
\end{figure*}

Supplementary Figure~\ref{fig:delay} shows the experimental transient evolution of the T-MOKE asymmetry in the marked energy regions. We note that the dynamics strongly differ for different incidence angles $\theta$.

Integrating over extended energy regions results in a "mixing" of the spectral features: at the smaller incidence angle, the integral of the ultrafast increase around 64~eV and the decrease around 66~eV is measured. This leads to the previously reported delay of the ultrafast dynamics of Ni compared to Fe. 
In contrast, the delay in the energy-integrated signal vanishes at larger incidence angles, because the increase at $h\nu = 64$~eV is not present in the transient asymmetry (cf. Fig.~2 in the main text).

Therefore, the transient evolution of the T-MOKE asymmetry is not always a good indication of the magnetic moment and depends strongly on the incidence angle, as described in the previous section and in detail in Ref.~\cite{probst_analysis_2023}. In consequence, we performed an element-specific evaluation of the off-diagonal tensor element \reexy{}, which is shown in Supplementary Figure~\ref{fig:delayexy}.

\begin{figure*}[tbh]
\centering
    \includegraphics[width=0.85\textwidth]{{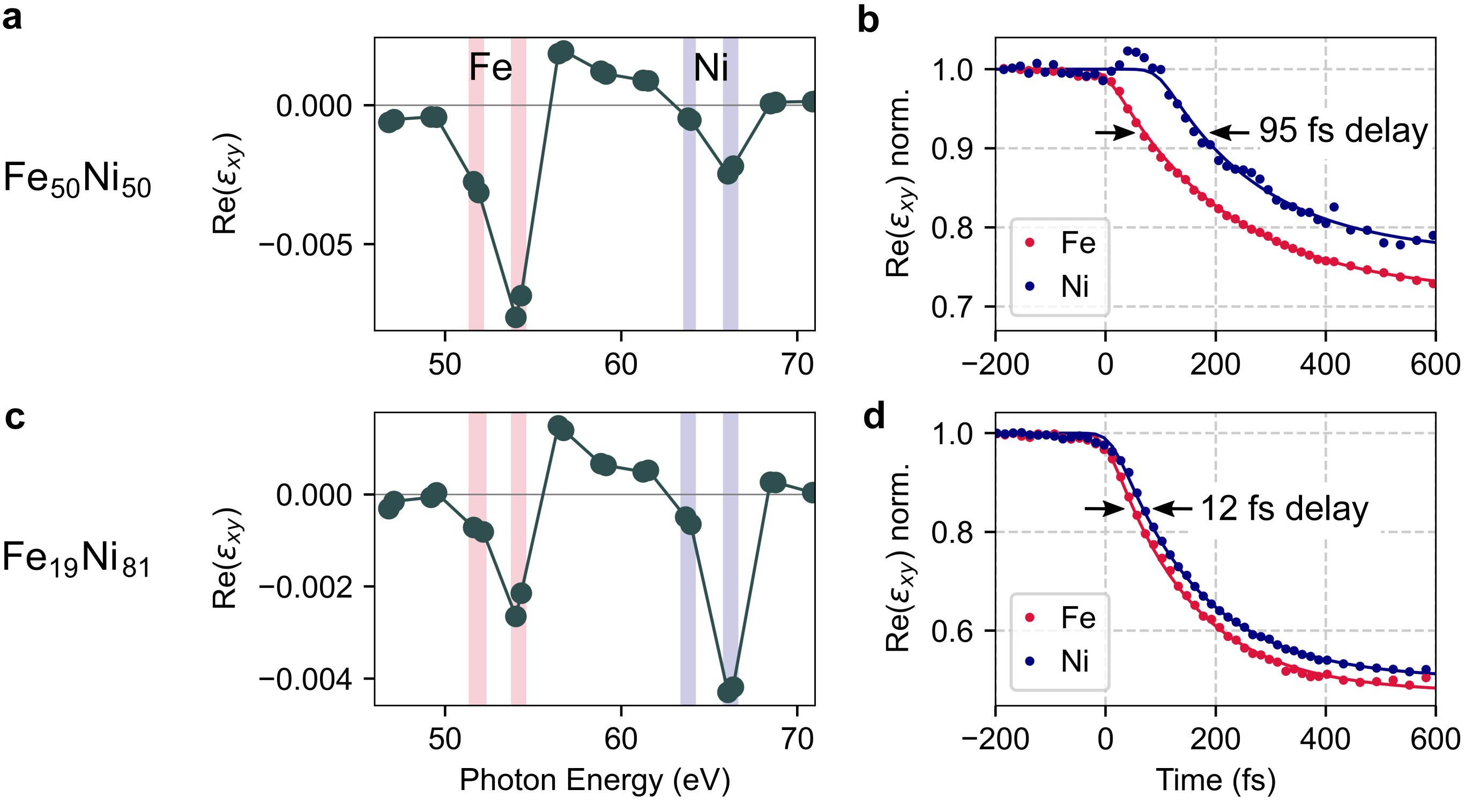}}
    \caption{a) Off-diagonal tensor element of \feni{} and c) \permalloy{}. b) and d) Transient evolution of \exy{} integrated for multiple harmonics around each absorption edge: Fe: 51.6, 51.9, 54.0, and 54.3~eV; Ni: 63.6, 63.9, 66.1 and 66.4~eV.}
    \label{fig:delayexy}
\end{figure*}

We fitted the transient evolution of \reexy{} with a simple exponential decay fit given by
\begin{equation}
		\frac{A(t)}{A(t<t_0)} = G(t) \circledast  \bigg[ 1-\Theta(t-t_0)\cdot 
		\Delta A_m \cdot \left( 1- e^{- \frac{t-t_0}{\tau_m}}\right)
	 \bigg].
	\label{time_dependent}
\end{equation}
Here, $t_0$ defines the onset of the dynamics, $\tau_m$ is the demagnetization constant, and $\Delta A_m$ is proportional to the maximum demagnetization. The fit results are shown in Table~\ref{tab:fitresults}.
We observe a delayed onset of the dynamics of \reexy{} for Ni compared to Fe that amounts to $12 \pm 3$~fs for \permalloy{} and to $95 \pm 7$~fs for \feni{}.

\begin{table}[tbh]
\begin{tabular}{cccc|cccc}
\multicolumn{4}{c|}{\feni{}}& \multicolumn{3}{c}{\permalloy{}}\\
	  & Fe & Ni&&&Fe&Ni\\\hline
	$t_ 0$~(fs)     & $ 0 \pm 2  $          & $ 95 \pm 6$ && & $ 0 \pm 2 $          & $ 12 \pm 2$            \\ 
	$\Delta A_m$    & $ 0.286 \pm 0.003 $   & $ 0.234 \pm 0.008 $ &&& $ 0.525 \pm 0.004 $   & $ 0.498 \pm 0.004 $   \\
	$\tau_m$~(fs)   & $ 215\pm 6 $         & $ 184 \pm 19$        &&& $ 147\pm 4 $         & $ 153 \pm 3$   \\ \hline
\end{tabular}
\label{tab:fitresults}
\centering
\caption{Fit results of the exponential fits to the transient evolution of \reexy{} (Supplementary Figure~\ref{fig:delayexy}) for Fe and Ni in both FeNi-alloys.} 
\end{table}

\newpage
\section{TDDFT: OISTR and spin flips in FeNi}
\begin{figure*}[htb!]
\centering
    \includegraphics[width=0.85\textwidth]{{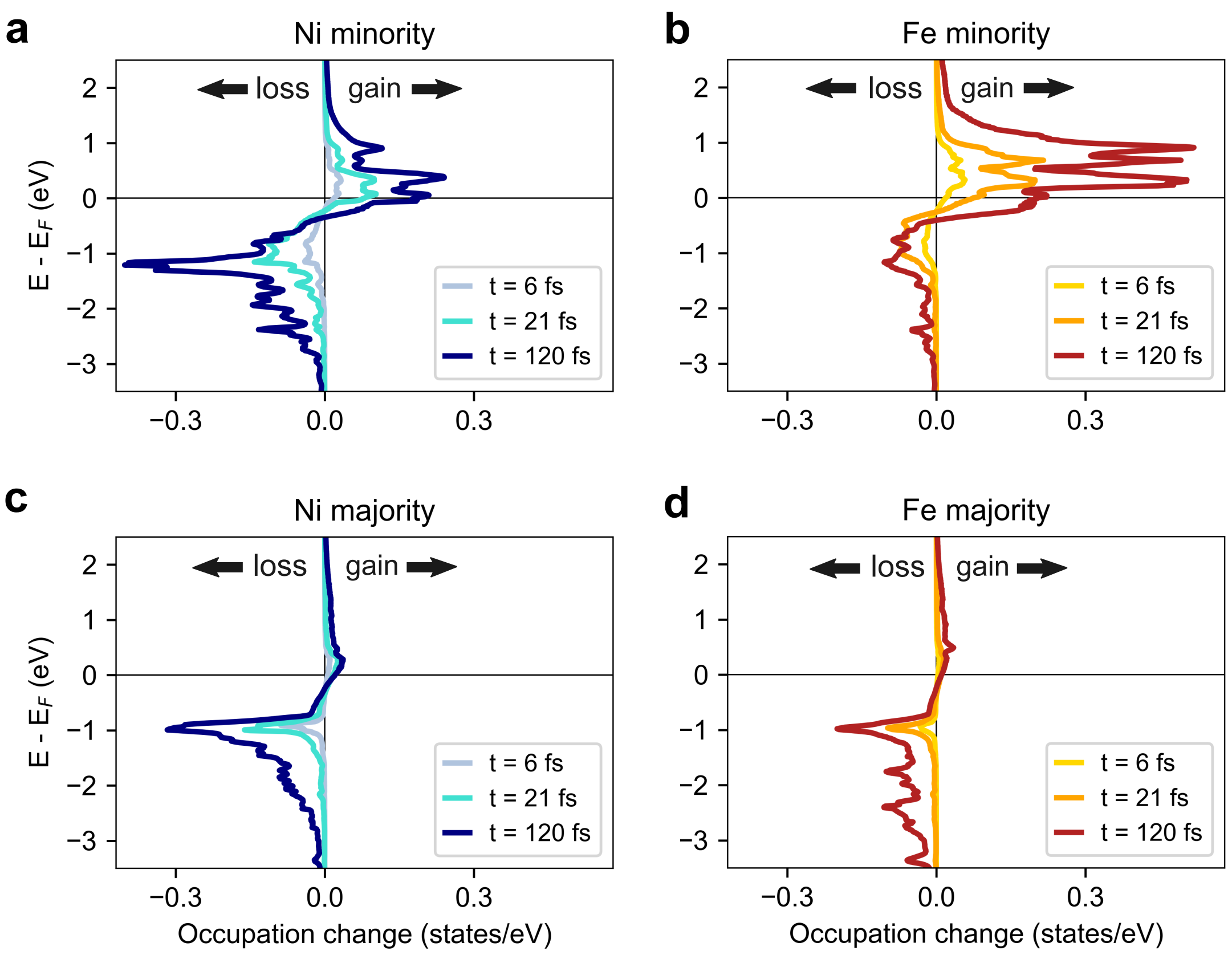}}
    \caption{TDDFT calculations of the transient occupation change in \feni{} for minority carriers (a) and (b) and majority carriers (c) and (d). 
   }
    \label{fig:TDDFT_DOS}
\end{figure*}

Fig.~\ref{fig:TDDFT_DOS} shows the transient changes in the occupation of the minority and majority channels in \feni{} calculated by TDDFT. As expected, optical excitation with a 1.2~eV pump pulse will induce a decrease of carriers below the Fermi level and an increase above the Fermi level. The incident fluence is 18~$\nicefrac{\text{mJ}}{\text{cm}^2}$.

The optical intersite spin transfer (OISTR) can be seen by the strong loss of minority Ni carriers around 1~eV below E$_F$ and by a strong increase of Fe minority carriers $\approx$~1~eV above E$_f$. This process happens directly at the onset of the optical laser pulse (47~fs FWHM). At a time of t=120~fs we find that scattering processes already dominate the dynamics. Spin-orbit mediated spin flips below the Fermi level lead to the strong losses in the majority channel both for Ni and Fe.

%% file: main_arxiv.bbl
\begin{thebibliography}{32}%
\makeatletter
\providecommand \@ifxundefined [1]{%
 \@ifx{#1\undefined}
}%
\providecommand \@ifnum [1]{%
 \ifnum #1\expandafter \@firstoftwo
 \else \expandafter \@secondoftwo
 \fi
}%
\providecommand \@ifx [1]{%
 \ifx #1\expandafter \@firstoftwo
 \else \expandafter \@secondoftwo
 \fi
}%
\providecommand \natexlab [1]{#1}%
\providecommand \enquote  [1]{``#1''}%
\providecommand \bibnamefont  [1]{#1}%
\providecommand \bibfnamefont [1]{#1}%
\providecommand \citenamefont [1]{#1}%
\providecommand \href@noop [0]{\@secondoftwo}%
\providecommand \href [0]{\begingroup \@sanitize@url \@href}%
\providecommand \@href[1]{\@@startlink{#1}\@@href}%
\providecommand \@@href[1]{\endgroup#1\@@endlink}%
\providecommand \@sanitize@url [0]{\catcode `\\12\catcode `\$12\catcode
  `\&12\catcode `\#12\catcode `\^12\catcode `\_12\catcode `\%12\relax}%
\providecommand \@@startlink[1]{}%
\providecommand \@@endlink[0]{}%
\providecommand \url  [0]{\begingroup\@sanitize@url \@url }%
\providecommand \@url [1]{\endgroup\@href {#1}{\urlprefix }}%
\providecommand \urlprefix  [0]{URL }%
\providecommand \Eprint [0]{\href }%
\providecommand \doibase [0]{http://dx.doi.org/}%
\providecommand \selectlanguage [0]{\@gobble}%
\providecommand \bibinfo  [0]{\@secondoftwo}%
\providecommand \bibfield  [0]{\@secondoftwo}%
\providecommand \translation [1]{[#1]}%
\providecommand \BibitemOpen [0]{}%
\providecommand \bibitemStop [0]{}%
\providecommand \bibitemNoStop [0]{.\EOS\space}%
\providecommand \EOS [0]{\spacefactor3000\relax}%
\providecommand \BibitemShut  [1]{\csname bibitem#1\endcsname}%
\let\auto@bib@innerbib\@empty
\bibitem [{\citenamefont {Malinowski}\ \emph {et~al.}(2008)\citenamefont
  {Malinowski}, \citenamefont {Longa}, \citenamefont {Rietjens}, \citenamefont
  {Paluskar}, \citenamefont {Huijink}, \citenamefont {Swagten},\ and\
  \citenamefont {Koopmans}}]{Malinowski:2008co}%
  \BibitemOpen
  \bibfield  {author} {\bibinfo {author} {\bibfnamefont {G.}~\bibnamefont
  {Malinowski}}, \bibinfo {author} {\bibfnamefont {F.~D.}\ \bibnamefont
  {Longa}}, \bibinfo {author} {\bibfnamefont {J.~H.~H.}\ \bibnamefont
  {Rietjens}}, \bibinfo {author} {\bibfnamefont {P.~V.}\ \bibnamefont
  {Paluskar}}, \bibinfo {author} {\bibfnamefont {R.}~\bibnamefont {Huijink}},
  \bibinfo {author} {\bibfnamefont {H.~J.~M.}\ \bibnamefont {Swagten}}, \ and\
  \bibinfo {author} {\bibfnamefont {B.}~\bibnamefont {Koopmans}},\ }\href@noop
  {} {\bibfield  {journal} {\bibinfo  {journal} {Nature Physics}\ }\textbf
  {\bibinfo {volume} {4}},\ \bibinfo {pages} {855} (\bibinfo {year}
  {2008})}\BibitemShut {NoStop}%
\bibitem [{\citenamefont {Rudolf}\ \emph {et~al.}(2012)\citenamefont {Rudolf},
  \citenamefont {La-O-Vorakiat}, \citenamefont {Battiato}, \citenamefont
  {Adam}, \citenamefont {Shaw}, \citenamefont {Turgut}, \citenamefont
  {Maldonado}, \citenamefont {Mathias}, \citenamefont {Grychtol}, \citenamefont
  {Nembach}, \citenamefont {Silva}, \citenamefont {Aeschlimann}, \citenamefont
  {Kapteyn}, \citenamefont {Murnane}, \citenamefont {Schneider},\ and\
  \citenamefont {Oppeneer}}]{rudolf_ultrafast_2012}%
  \BibitemOpen
  \bibfield  {author} {\bibinfo {author} {\bibfnamefont {D.}~\bibnamefont
  {Rudolf}}, \bibinfo {author} {\bibfnamefont {C.}~\bibnamefont
  {La-O-Vorakiat}}, \bibinfo {author} {\bibfnamefont {M.}~\bibnamefont
  {Battiato}}, \bibinfo {author} {\bibfnamefont {R.}~\bibnamefont {Adam}},
  \bibinfo {author} {\bibfnamefont {J.~M.}\ \bibnamefont {Shaw}}, \bibinfo
  {author} {\bibfnamefont {E.}~\bibnamefont {Turgut}}, \bibinfo {author}
  {\bibfnamefont {P.}~\bibnamefont {Maldonado}}, \bibinfo {author}
  {\bibfnamefont {S.}~\bibnamefont {Mathias}}, \bibinfo {author} {\bibfnamefont
  {P.}~\bibnamefont {Grychtol}}, \bibinfo {author} {\bibfnamefont {H.~T.}\
  \bibnamefont {Nembach}}, \bibinfo {author} {\bibfnamefont {T.~J.}\
  \bibnamefont {Silva}}, \bibinfo {author} {\bibfnamefont {M.}~\bibnamefont
  {Aeschlimann}}, \bibinfo {author} {\bibfnamefont {H.~C.}\ \bibnamefont
  {Kapteyn}}, \bibinfo {author} {\bibfnamefont {M.~M.}\ \bibnamefont
  {Murnane}}, \bibinfo {author} {\bibfnamefont {C.~M.}\ \bibnamefont
  {Schneider}}, \ and\ \bibinfo {author} {\bibfnamefont {P.~M.}\ \bibnamefont
  {Oppeneer}},\ }\href {\doibase 10.1038/ncomms2029} {\bibfield  {journal}
  {\bibinfo  {journal} {Nat Commun}\ }\textbf {\bibinfo {volume} {3}},\
  \bibinfo {pages} {1037} (\bibinfo {year} {2012})}\BibitemShut {NoStop}%
\bibitem [{\citenamefont {Battiato}\ \emph {et~al.}(2010)\citenamefont
  {Battiato}, \citenamefont {Carva},\ and\ \citenamefont
  {Oppeneer}}]{battiato_superdiffusive_2010}%
  \BibitemOpen
  \bibfield  {author} {\bibinfo {author} {\bibfnamefont {M.}~\bibnamefont
  {Battiato}}, \bibinfo {author} {\bibfnamefont {K.}~\bibnamefont {Carva}}, \
  and\ \bibinfo {author} {\bibfnamefont {P.~M.}\ \bibnamefont {Oppeneer}},\
  }\href {\doibase 10.1103/PhysRevLett.105.027203} {\bibfield  {journal}
  {\bibinfo  {journal} {Physical Review Letters}\ }\textbf {\bibinfo {volume}
  {105}},\ \bibinfo {pages} {027203} (\bibinfo {year} {2010})}\BibitemShut
  {NoStop}%
\bibitem [{\citenamefont {Melnikov}\ \emph {et~al.}(2011)\citenamefont
  {Melnikov}, \citenamefont {Razdolski}, \citenamefont {Wehling}, \citenamefont
  {Papaioannou}, \citenamefont {Roddatis}, \citenamefont {Fumagalli},
  \citenamefont {Aktsipetrov}, \citenamefont {Lichtenstein},\ and\
  \citenamefont {Bovensiepen}}]{melnikov_ultrafast_2011}%
  \BibitemOpen
  \bibfield  {author} {\bibinfo {author} {\bibfnamefont {A.}~\bibnamefont
  {Melnikov}}, \bibinfo {author} {\bibfnamefont {I.}~\bibnamefont {Razdolski}},
  \bibinfo {author} {\bibfnamefont {T.~O.}\ \bibnamefont {Wehling}}, \bibinfo
  {author} {\bibfnamefont {E.~T.}\ \bibnamefont {Papaioannou}}, \bibinfo
  {author} {\bibfnamefont {V.}~\bibnamefont {Roddatis}}, \bibinfo {author}
  {\bibfnamefont {P.}~\bibnamefont {Fumagalli}}, \bibinfo {author}
  {\bibfnamefont {O.}~\bibnamefont {Aktsipetrov}}, \bibinfo {author}
  {\bibfnamefont {A.~I.}\ \bibnamefont {Lichtenstein}}, \ and\ \bibinfo
  {author} {\bibfnamefont {U.}~\bibnamefont {Bovensiepen}},\ }\href {\doibase
  10.1103/PhysRevLett.107.076601} {\bibfield  {journal} {\bibinfo  {journal}
  {Physical Review Letters}\ }\textbf {\bibinfo {volume} {107}},\ \bibinfo
  {pages} {076601} (\bibinfo {year} {2011})}\BibitemShut {NoStop}%
\bibitem [{\citenamefont {Seifert}\ \emph {et~al.}(2016)\citenamefont
  {Seifert}, \citenamefont {Jaiswal}, \citenamefont {Martens}, \citenamefont
  {Hannegan}, \citenamefont {Braun}, \citenamefont {Maldonado}, \citenamefont
  {Freimuth}, \citenamefont {Kronenberg}, \citenamefont {Henrizi},
  \citenamefont {Radu}, \citenamefont {Beaurepaire}, \citenamefont {Mokrousov},
  \citenamefont {Oppeneer}, \citenamefont {Jourdan}, \citenamefont {Jakob},
  \citenamefont {Turchinovich}, \citenamefont {Hayden}, \citenamefont {Wolf},
  \citenamefont {Münzenberg}, \citenamefont {Kläui},\ and\ \citenamefont
  {Kampfrath}}]{seifert_efficient_2016}%
  \BibitemOpen
  \bibfield  {author} {\bibinfo {author} {\bibfnamefont {T.}~\bibnamefont
  {Seifert}}, \bibinfo {author} {\bibfnamefont {S.}~\bibnamefont {Jaiswal}},
  \bibinfo {author} {\bibfnamefont {U.}~\bibnamefont {Martens}}, \bibinfo
  {author} {\bibfnamefont {J.}~\bibnamefont {Hannegan}}, \bibinfo {author}
  {\bibfnamefont {L.}~\bibnamefont {Braun}}, \bibinfo {author} {\bibfnamefont
  {P.}~\bibnamefont {Maldonado}}, \bibinfo {author} {\bibfnamefont
  {F.}~\bibnamefont {Freimuth}}, \bibinfo {author} {\bibfnamefont
  {A.}~\bibnamefont {Kronenberg}}, \bibinfo {author} {\bibfnamefont
  {J.}~\bibnamefont {Henrizi}}, \bibinfo {author} {\bibfnamefont
  {I.}~\bibnamefont {Radu}}, \bibinfo {author} {\bibfnamefont {E.}~\bibnamefont
  {Beaurepaire}}, \bibinfo {author} {\bibfnamefont {Y.}~\bibnamefont
  {Mokrousov}}, \bibinfo {author} {\bibfnamefont {P.~M.}\ \bibnamefont
  {Oppeneer}}, \bibinfo {author} {\bibfnamefont {M.}~\bibnamefont {Jourdan}},
  \bibinfo {author} {\bibfnamefont {G.}~\bibnamefont {Jakob}}, \bibinfo
  {author} {\bibfnamefont {D.}~\bibnamefont {Turchinovich}}, \bibinfo {author}
  {\bibfnamefont {L.~M.}\ \bibnamefont {Hayden}}, \bibinfo {author}
  {\bibfnamefont {M.}~\bibnamefont {Wolf}}, \bibinfo {author} {\bibfnamefont
  {M.}~\bibnamefont {Münzenberg}}, \bibinfo {author} {\bibfnamefont
  {M.}~\bibnamefont {Kläui}}, \ and\ \bibinfo {author} {\bibfnamefont
  {T.}~\bibnamefont {Kampfrath}},\ }\href {\doibase 10.1038/nphoton.2016.91}
  {\bibfield  {journal} {\bibinfo  {journal} {Nature Photon}\ }\textbf
  {\bibinfo {volume} {10}},\ \bibinfo {pages} {483} (\bibinfo {year}
  {2016})}\BibitemShut {NoStop}%
\bibitem [{\citenamefont {Stanciu}\ \emph {et~al.}(2007)\citenamefont
  {Stanciu}, \citenamefont {Hansteen}, \citenamefont {Kimel}, \citenamefont
  {Kirilyuk}, \citenamefont {Tsukamoto}, \citenamefont {Itoh},\ and\
  \citenamefont {Rasing}}]{Stanciu:2007fy}%
  \BibitemOpen
  \bibfield  {author} {\bibinfo {author} {\bibfnamefont {C.~D.}\ \bibnamefont
  {Stanciu}}, \bibinfo {author} {\bibfnamefont {F.}~\bibnamefont {Hansteen}},
  \bibinfo {author} {\bibfnamefont {A.~V.}\ \bibnamefont {Kimel}}, \bibinfo
  {author} {\bibfnamefont {A.}~\bibnamefont {Kirilyuk}}, \bibinfo {author}
  {\bibfnamefont {A.}~\bibnamefont {Tsukamoto}}, \bibinfo {author}
  {\bibfnamefont {A.}~\bibnamefont {Itoh}}, \ and\ \bibinfo {author}
  {\bibfnamefont {T.}~\bibnamefont {Rasing}},\ }\href {\doibase
  10.1103/PhysRevLett.99.047601} {\bibfield  {journal} {\bibinfo  {journal}
  {Phys. Rev. Lett.}\ }\textbf {\bibinfo {volume} {99}},\ \bibinfo {pages}
  {047601} (\bibinfo {year} {2007})}\BibitemShut {NoStop}%
\bibitem [{\citenamefont {Lambert}\ \emph {et~al.}(2014)\citenamefont
  {Lambert}, \citenamefont {Mangin}, \citenamefont {Varaprasad}, \citenamefont
  {Takahashi}, \citenamefont {Hehn}, \citenamefont {Cinchetti}, \citenamefont
  {Malinowski}, \citenamefont {Hono}, \citenamefont {Fainman}, \citenamefont
  {Aeschlimann},\ and\ \citenamefont {Fullerton}}]{Lambert_Science_2014}%
  \BibitemOpen
  \bibfield  {author} {\bibinfo {author} {\bibfnamefont {C.-H.}\ \bibnamefont
  {Lambert}}, \bibinfo {author} {\bibfnamefont {S.}~\bibnamefont {Mangin}},
  \bibinfo {author} {\bibfnamefont {C.~B.}\ \bibnamefont {Varaprasad}},
  \bibinfo {author} {\bibfnamefont {Y.}~\bibnamefont {Takahashi}}, \bibinfo
  {author} {\bibfnamefont {M.}~\bibnamefont {Hehn}}, \bibinfo {author}
  {\bibfnamefont {M.}~\bibnamefont {Cinchetti}}, \bibinfo {author}
  {\bibfnamefont {G.}~\bibnamefont {Malinowski}}, \bibinfo {author}
  {\bibfnamefont {K.}~\bibnamefont {Hono}}, \bibinfo {author} {\bibfnamefont
  {Y.}~\bibnamefont {Fainman}}, \bibinfo {author} {\bibfnamefont
  {M.}~\bibnamefont {Aeschlimann}}, \ and\ \bibinfo {author} {\bibfnamefont
  {E.~E.}\ \bibnamefont {Fullerton}},\ }\href {\doibase
  10.1126/science.1253493} {\bibfield  {journal} {\bibinfo  {journal}
  {Science}\ }\textbf {\bibinfo {volume} {345}},\ \bibinfo {pages} {1337}
  (\bibinfo {year} {2014})},\ \Eprint {http://arxiv.org/abs/1403.0784}
  {1403.0784} \BibitemShut {NoStop}%
\bibitem [{\citenamefont {Schlauderer}\ \emph {et~al.}(2019)\citenamefont
  {Schlauderer}, \citenamefont {Lange}, \citenamefont {Baierl}, \citenamefont
  {Ebnet}, \citenamefont {Schmid}, \citenamefont {Valovcin}, \citenamefont
  {Zvezdin}, \citenamefont {Kimel}, \citenamefont {Mikhaylovskiy},\ and\
  \citenamefont {Huber}}]{schlauderer_temporal_2019}%
  \BibitemOpen
  \bibfield  {author} {\bibinfo {author} {\bibfnamefont {S.}~\bibnamefont
  {Schlauderer}}, \bibinfo {author} {\bibfnamefont {C.}~\bibnamefont {Lange}},
  \bibinfo {author} {\bibfnamefont {S.}~\bibnamefont {Baierl}}, \bibinfo
  {author} {\bibfnamefont {T.}~\bibnamefont {Ebnet}}, \bibinfo {author}
  {\bibfnamefont {C.~P.}\ \bibnamefont {Schmid}}, \bibinfo {author}
  {\bibfnamefont {D.~C.}\ \bibnamefont {Valovcin}}, \bibinfo {author}
  {\bibfnamefont {A.~K.}\ \bibnamefont {Zvezdin}}, \bibinfo {author}
  {\bibfnamefont {A.~V.}\ \bibnamefont {Kimel}}, \bibinfo {author}
  {\bibfnamefont {R.~V.}\ \bibnamefont {Mikhaylovskiy}}, \ and\ \bibinfo
  {author} {\bibfnamefont {R.}~\bibnamefont {Huber}},\ }\href {\doibase
  10.1038/s41586-019-1174-7} {\bibfield  {journal} {\bibinfo  {journal}
  {Nature}\ }\textbf {\bibinfo {volume} {569}},\ \bibinfo {pages} {383}
  (\bibinfo {year} {2019})}\BibitemShut {NoStop}%
\bibitem [{\citenamefont {Hofherr}\ \emph {et~al.}(2020)\citenamefont
  {Hofherr}, \citenamefont {Häuser}, \citenamefont {Dewhurst}, \citenamefont
  {Tengdin}, \citenamefont {Sakshath}, \citenamefont {Nembach}, \citenamefont
  {Weber}, \citenamefont {Shaw}, \citenamefont {Silva}, \citenamefont
  {Kapteyn}, \citenamefont {Cinchetti}, \citenamefont {Rethfeld}, \citenamefont
  {Murnane}, \citenamefont {Steil}, \citenamefont {Stadtmüller}, \citenamefont
  {Sharma}, \citenamefont {Aeschlimann},\ and\ \citenamefont
  {Mathias}}]{hofherr_ultrafast_2020}%
  \BibitemOpen
  \bibfield  {author} {\bibinfo {author} {\bibfnamefont {M.}~\bibnamefont
  {Hofherr}}, \bibinfo {author} {\bibfnamefont {S.}~\bibnamefont {Häuser}},
  \bibinfo {author} {\bibfnamefont {J.~K.}\ \bibnamefont {Dewhurst}}, \bibinfo
  {author} {\bibfnamefont {P.}~\bibnamefont {Tengdin}}, \bibinfo {author}
  {\bibfnamefont {S.}~\bibnamefont {Sakshath}}, \bibinfo {author}
  {\bibfnamefont {H.~T.}\ \bibnamefont {Nembach}}, \bibinfo {author}
  {\bibfnamefont {S.~T.}\ \bibnamefont {Weber}}, \bibinfo {author}
  {\bibfnamefont {J.~M.}\ \bibnamefont {Shaw}}, \bibinfo {author}
  {\bibfnamefont {T.~J.}\ \bibnamefont {Silva}}, \bibinfo {author}
  {\bibfnamefont {H.~C.}\ \bibnamefont {Kapteyn}}, \bibinfo {author}
  {\bibfnamefont {M.}~\bibnamefont {Cinchetti}}, \bibinfo {author}
  {\bibfnamefont {B.}~\bibnamefont {Rethfeld}}, \bibinfo {author}
  {\bibfnamefont {M.~M.}\ \bibnamefont {Murnane}}, \bibinfo {author}
  {\bibfnamefont {D.}~\bibnamefont {Steil}}, \bibinfo {author} {\bibfnamefont
  {B.}~\bibnamefont {Stadtmüller}}, \bibinfo {author} {\bibfnamefont
  {S.}~\bibnamefont {Sharma}}, \bibinfo {author} {\bibfnamefont
  {M.}~\bibnamefont {Aeschlimann}}, \ and\ \bibinfo {author} {\bibfnamefont
  {S.}~\bibnamefont {Mathias}},\ }\href {\doibase 10.1126/sciadv.aay8717}
  {\bibfield  {journal} {\bibinfo  {journal} {Science Advances}\ }\textbf
  {\bibinfo {volume} {6}},\ \bibinfo {pages} {eaay8717} (\bibinfo {year}
  {2020})}\BibitemShut {NoStop}%
\bibitem [{\citenamefont {Tengdin}\ \emph {et~al.}(2020)\citenamefont
  {Tengdin}, \citenamefont {Gentry}, \citenamefont {Blonsky}, \citenamefont
  {Zusin}, \citenamefont {Gerrity}, \citenamefont {Hellbrück}, \citenamefont
  {Hofherr}, \citenamefont {Shaw}, \citenamefont {Kvashnin}, \citenamefont
  {Delczeg-Czirjak}, \citenamefont {Arora}, \citenamefont {Nembach},
  \citenamefont {Silva}, \citenamefont {Mathias}, \citenamefont {Aeschlimann},
  \citenamefont {Kapteyn}, \citenamefont {Thonig}, \citenamefont {Koumpouras},
  \citenamefont {Eriksson},\ and\ \citenamefont
  {Murnane}}]{tengdin_direct_2020}%
  \BibitemOpen
  \bibfield  {author} {\bibinfo {author} {\bibfnamefont {P.}~\bibnamefont
  {Tengdin}}, \bibinfo {author} {\bibfnamefont {C.}~\bibnamefont {Gentry}},
  \bibinfo {author} {\bibfnamefont {A.}~\bibnamefont {Blonsky}}, \bibinfo
  {author} {\bibfnamefont {D.}~\bibnamefont {Zusin}}, \bibinfo {author}
  {\bibfnamefont {M.}~\bibnamefont {Gerrity}}, \bibinfo {author} {\bibfnamefont
  {L.}~\bibnamefont {Hellbrück}}, \bibinfo {author} {\bibfnamefont
  {M.}~\bibnamefont {Hofherr}}, \bibinfo {author} {\bibfnamefont
  {J.}~\bibnamefont {Shaw}}, \bibinfo {author} {\bibfnamefont {Y.}~\bibnamefont
  {Kvashnin}}, \bibinfo {author} {\bibfnamefont {E.~K.}\ \bibnamefont
  {Delczeg-Czirjak}}, \bibinfo {author} {\bibfnamefont {M.}~\bibnamefont
  {Arora}}, \bibinfo {author} {\bibfnamefont {H.}~\bibnamefont {Nembach}},
  \bibinfo {author} {\bibfnamefont {T.~J.}\ \bibnamefont {Silva}}, \bibinfo
  {author} {\bibfnamefont {S.}~\bibnamefont {Mathias}}, \bibinfo {author}
  {\bibfnamefont {M.}~\bibnamefont {Aeschlimann}}, \bibinfo {author}
  {\bibfnamefont {H.~C.}\ \bibnamefont {Kapteyn}}, \bibinfo {author}
  {\bibfnamefont {D.}~\bibnamefont {Thonig}}, \bibinfo {author} {\bibfnamefont
  {K.}~\bibnamefont {Koumpouras}}, \bibinfo {author} {\bibfnamefont
  {O.}~\bibnamefont {Eriksson}}, \ and\ \bibinfo {author} {\bibfnamefont
  {M.~M.}\ \bibnamefont {Murnane}},\ }\href {\doibase 10.1126/sciadv.aaz1100}
  {\bibfield  {journal} {\bibinfo  {journal} {Science Advances}\ }\textbf
  {\bibinfo {volume} {6}},\ \bibinfo {pages} {eaaz1100} (\bibinfo {year}
  {2020})}\BibitemShut {NoStop}%
\bibitem [{\citenamefont {Willems}\ \emph {et~al.}(2020)\citenamefont
  {Willems}, \citenamefont {von Korff~Schmising}, \citenamefont {Strüber},
  \citenamefont {Schick}, \citenamefont {Engel}, \citenamefont {Dewhurst},
  \citenamefont {Elliott}, \citenamefont {Sharma},\ and\ \citenamefont
  {Eisebitt}}]{willems_optical_2020}%
  \BibitemOpen
  \bibfield  {author} {\bibinfo {author} {\bibfnamefont {F.}~\bibnamefont
  {Willems}}, \bibinfo {author} {\bibfnamefont {C.}~\bibnamefont {von
  Korff~Schmising}}, \bibinfo {author} {\bibfnamefont {C.}~\bibnamefont
  {Strüber}}, \bibinfo {author} {\bibfnamefont {D.}~\bibnamefont {Schick}},
  \bibinfo {author} {\bibfnamefont {D.~W.}\ \bibnamefont {Engel}}, \bibinfo
  {author} {\bibfnamefont {J.~K.}\ \bibnamefont {Dewhurst}}, \bibinfo {author}
  {\bibfnamefont {P.}~\bibnamefont {Elliott}}, \bibinfo {author} {\bibfnamefont
  {S.}~\bibnamefont {Sharma}}, \ and\ \bibinfo {author} {\bibfnamefont
  {S.}~\bibnamefont {Eisebitt}},\ }\href {\doibase 10.1038/s41467-020-14691-5}
  {\bibfield  {journal} {\bibinfo  {journal} {Nat Commun}\ }\textbf {\bibinfo
  {volume} {11}},\ \bibinfo {pages} {871} (\bibinfo {year} {2020})}\BibitemShut
  {NoStop}%
\bibitem [{\citenamefont {Siegrist}\ \emph {et~al.}(2019)\citenamefont
  {Siegrist}, \citenamefont {Gessner}, \citenamefont {Ossiander}, \citenamefont
  {Denker}, \citenamefont {Chang}, \citenamefont {Schröder}, \citenamefont
  {Guggenmos}, \citenamefont {Cui}, \citenamefont {Walowski}, \citenamefont
  {Martens}, \citenamefont {Dewhurst}, \citenamefont {Kleineberg},
  \citenamefont {Münzenberg}, \citenamefont {Sharma},\ and\ \citenamefont
  {Schultze}}]{siegrist_light-wave_2019}%
  \BibitemOpen
  \bibfield  {author} {\bibinfo {author} {\bibfnamefont {F.}~\bibnamefont
  {Siegrist}}, \bibinfo {author} {\bibfnamefont {J.~A.}\ \bibnamefont
  {Gessner}}, \bibinfo {author} {\bibfnamefont {M.}~\bibnamefont {Ossiander}},
  \bibinfo {author} {\bibfnamefont {C.}~\bibnamefont {Denker}}, \bibinfo
  {author} {\bibfnamefont {Y.-P.}\ \bibnamefont {Chang}}, \bibinfo {author}
  {\bibfnamefont {M.~C.}\ \bibnamefont {Schröder}}, \bibinfo {author}
  {\bibfnamefont {A.}~\bibnamefont {Guggenmos}}, \bibinfo {author}
  {\bibfnamefont {Y.}~\bibnamefont {Cui}}, \bibinfo {author} {\bibfnamefont
  {J.}~\bibnamefont {Walowski}}, \bibinfo {author} {\bibfnamefont
  {U.}~\bibnamefont {Martens}}, \bibinfo {author} {\bibfnamefont {J.~K.}\
  \bibnamefont {Dewhurst}}, \bibinfo {author} {\bibfnamefont {U.}~\bibnamefont
  {Kleineberg}}, \bibinfo {author} {\bibfnamefont {M.}~\bibnamefont
  {Münzenberg}}, \bibinfo {author} {\bibfnamefont {S.}~\bibnamefont {Sharma}},
  \ and\ \bibinfo {author} {\bibfnamefont {M.}~\bibnamefont {Schultze}},\
  }\href {\doibase 10.1038/s41586-019-1333-x} {\bibfield  {journal} {\bibinfo
  {journal} {Nature}\ }\textbf {\bibinfo {volume} {571}},\ \bibinfo {pages}
  {240} (\bibinfo {year} {2019})}\BibitemShut {NoStop}%
\bibitem [{\citenamefont {Steil}\ \emph {et~al.}(2020)\citenamefont {Steil},
  \citenamefont {Walowski}, \citenamefont {Gerhard}, \citenamefont {Kiessling},
  \citenamefont {Ebke}, \citenamefont {Thomas}, \citenamefont {Kubota},
  \citenamefont {Oogane}, \citenamefont {Ando}, \citenamefont {Otto},
  \citenamefont {Mann}, \citenamefont {Hofherr}, \citenamefont {Elliott},
  \citenamefont {Dewhurst}, \citenamefont {Reiss}, \citenamefont {Molenkamp},
  \citenamefont {Aeschlimann}, \citenamefont {Cinchetti}, \citenamefont
  {Münzenberg}, \citenamefont {Sharma},\ and\ \citenamefont
  {Mathias}}]{steil_efficiency_2020}%
  \BibitemOpen
  \bibfield  {author} {\bibinfo {author} {\bibfnamefont {D.}~\bibnamefont
  {Steil}}, \bibinfo {author} {\bibfnamefont {J.}~\bibnamefont {Walowski}},
  \bibinfo {author} {\bibfnamefont {F.}~\bibnamefont {Gerhard}}, \bibinfo
  {author} {\bibfnamefont {T.}~\bibnamefont {Kiessling}}, \bibinfo {author}
  {\bibfnamefont {D.}~\bibnamefont {Ebke}}, \bibinfo {author} {\bibfnamefont
  {A.}~\bibnamefont {Thomas}}, \bibinfo {author} {\bibfnamefont
  {T.}~\bibnamefont {Kubota}}, \bibinfo {author} {\bibfnamefont
  {M.}~\bibnamefont {Oogane}}, \bibinfo {author} {\bibfnamefont
  {Y.}~\bibnamefont {Ando}}, \bibinfo {author} {\bibfnamefont {J.}~\bibnamefont
  {Otto}}, \bibinfo {author} {\bibfnamefont {A.}~\bibnamefont {Mann}}, \bibinfo
  {author} {\bibfnamefont {M.}~\bibnamefont {Hofherr}}, \bibinfo {author}
  {\bibfnamefont {P.}~\bibnamefont {Elliott}}, \bibinfo {author} {\bibfnamefont
  {J.~K.}\ \bibnamefont {Dewhurst}}, \bibinfo {author} {\bibfnamefont
  {G.}~\bibnamefont {Reiss}}, \bibinfo {author} {\bibfnamefont
  {L.}~\bibnamefont {Molenkamp}}, \bibinfo {author} {\bibfnamefont
  {M.}~\bibnamefont {Aeschlimann}}, \bibinfo {author} {\bibfnamefont
  {M.}~\bibnamefont {Cinchetti}}, \bibinfo {author} {\bibfnamefont
  {M.}~\bibnamefont {Münzenberg}}, \bibinfo {author} {\bibfnamefont
  {S.}~\bibnamefont {Sharma}}, \ and\ \bibinfo {author} {\bibfnamefont
  {S.}~\bibnamefont {Mathias}},\ }\href {\doibase
  10.1103/PhysRevResearch.2.023199} {\bibfield  {journal} {\bibinfo  {journal}
  {Phys. Rev. Research}\ }\textbf {\bibinfo {volume} {2}},\ \bibinfo {pages}
  {023199} (\bibinfo {year} {2020})}\BibitemShut {NoStop}%
\bibitem [{\citenamefont {Ryan}\ \emph {et~al.}(2023)\citenamefont {Ryan},
  \citenamefont {Johnsen}, \citenamefont {Elhanoty}, \citenamefont {Grafov},
  \citenamefont {Li}, \citenamefont {Delin}, \citenamefont {Markou},
  \citenamefont {Lesne}, \citenamefont {Felser}, \citenamefont {Eriksson},
  \citenamefont {Kapteyn}, \citenamefont {Grånäs},\ and\ \citenamefont
  {Murnane}}]{Ryan.2023}%
  \BibitemOpen
  \bibfield  {author} {\bibinfo {author} {\bibfnamefont {S.~A.}\ \bibnamefont
  {Ryan}}, \bibinfo {author} {\bibfnamefont {P.~C.}\ \bibnamefont {Johnsen}},
  \bibinfo {author} {\bibfnamefont {M.~F.}\ \bibnamefont {Elhanoty}}, \bibinfo
  {author} {\bibfnamefont {A.}~\bibnamefont {Grafov}}, \bibinfo {author}
  {\bibfnamefont {N.}~\bibnamefont {Li}}, \bibinfo {author} {\bibfnamefont
  {A.}~\bibnamefont {Delin}}, \bibinfo {author} {\bibfnamefont
  {A.}~\bibnamefont {Markou}}, \bibinfo {author} {\bibfnamefont
  {E.}~\bibnamefont {Lesne}}, \bibinfo {author} {\bibfnamefont
  {C.}~\bibnamefont {Felser}}, \bibinfo {author} {\bibfnamefont
  {O.}~\bibnamefont {Eriksson}}, \bibinfo {author} {\bibfnamefont {H.~C.}\
  \bibnamefont {Kapteyn}}, \bibinfo {author} {\bibfnamefont {O.}~\bibnamefont
  {Grånäs}}, \ and\ \bibinfo {author} {\bibfnamefont {M.~M.}\ \bibnamefont
  {Murnane}},\ }\href@noop {} {\bibfield  {journal} {\bibinfo  {journal}
  {arXiv}\ } (\bibinfo {year} {2023})},\ \Eprint
  {http://arxiv.org/abs/2305.16455} {2305.16455} \BibitemShut {NoStop}%
\bibitem [{\citenamefont {El-Ghazaly}\ \emph {et~al.}(2020)\citenamefont
  {El-Ghazaly}, \citenamefont {Gorchon}, \citenamefont {Wilson}, \citenamefont
  {Pattabi},\ and\ \citenamefont {Bokor}}]{el-ghazaly_progress_2020}%
  \BibitemOpen
  \bibfield  {author} {\bibinfo {author} {\bibfnamefont {A.}~\bibnamefont
  {El-Ghazaly}}, \bibinfo {author} {\bibfnamefont {J.}~\bibnamefont {Gorchon}},
  \bibinfo {author} {\bibfnamefont {R.~B.}\ \bibnamefont {Wilson}}, \bibinfo
  {author} {\bibfnamefont {A.}~\bibnamefont {Pattabi}}, \ and\ \bibinfo
  {author} {\bibfnamefont {J.}~\bibnamefont {Bokor}},\ }\href {\doibase
  10.1016/j.jmmm.2020.166478} {\bibfield  {journal} {\bibinfo  {journal}
  {Journal of Magnetism and Magnetic Materials}\ }\textbf {\bibinfo {volume}
  {502}},\ \bibinfo {pages} {166478} (\bibinfo {year} {2020})}\BibitemShut
  {NoStop}%
\bibitem [{\citenamefont {Dewhurst}\ \emph {et~al.}(2018)\citenamefont
  {Dewhurst}, \citenamefont {Elliott}, \citenamefont {Shallcross},
  \citenamefont {Gross},\ and\ \citenamefont {Sharma}}]{Dewhurst2018}%
  \BibitemOpen
  \bibfield  {author} {\bibinfo {author} {\bibfnamefont {J.~K.}\ \bibnamefont
  {Dewhurst}}, \bibinfo {author} {\bibfnamefont {P.}~\bibnamefont {Elliott}},
  \bibinfo {author} {\bibfnamefont {S.}~\bibnamefont {Shallcross}}, \bibinfo
  {author} {\bibfnamefont {E.~K.~U.}\ \bibnamefont {Gross}}, \ and\ \bibinfo
  {author} {\bibfnamefont {S.}~\bibnamefont {Sharma}},\ }\href {\doibase
  10.1021/acs.nanolett.7b05118} {\bibfield  {journal} {\bibinfo  {journal}
  {Nano Letters}\ }\textbf {\bibinfo {volume} {18}},\ \bibinfo {pages} {1842}
  (\bibinfo {year} {2018})}\BibitemShut {NoStop}%
\bibitem [{\citenamefont {Minár}\ \emph {et~al.}(2014)\citenamefont {Minár},
  \citenamefont {Mankovsky}, \citenamefont {Šipr}, \citenamefont {Benea},\
  and\ \citenamefont {Ebert}}]{Minar_2014}%
  \BibitemOpen
  \bibfield  {author} {\bibinfo {author} {\bibfnamefont {J.}~\bibnamefont
  {Minár}}, \bibinfo {author} {\bibfnamefont {S.}~\bibnamefont {Mankovsky}},
  \bibinfo {author} {\bibfnamefont {O.}~\bibnamefont {Šipr}}, \bibinfo
  {author} {\bibfnamefont {D.}~\bibnamefont {Benea}}, \ and\ \bibinfo {author}
  {\bibfnamefont {H.}~\bibnamefont {Ebert}},\ }\href {\doibase
  10.1088/0953-8984/26/27/274206} {\bibfield  {journal} {\bibinfo  {journal}
  {Journal of Physics: Condensed Matter}\ }\textbf {\bibinfo {volume} {26}},\
  \bibinfo {pages} {274206} (\bibinfo {year} {2014})}\BibitemShut {NoStop}%
\bibitem [{\citenamefont {Hennes}\ \emph {et~al.}(2021)\citenamefont {Hennes},
  \citenamefont {Rösner}, \citenamefont {Chardonnet}, \citenamefont
  {Chiuzbaian}, \citenamefont {Delaunay}, \citenamefont {Döring},
  \citenamefont {Guzenko}, \citenamefont {Hehn}, \citenamefont {Jarrier},
  \citenamefont {Kleibert}, \citenamefont {Lebugle}, \citenamefont {Lüning},
  \citenamefont {Malinowski}, \citenamefont {Merhe}, \citenamefont {Naumenko},
  \citenamefont {Nikolov}, \citenamefont {Lopez-Quintas}, \citenamefont
  {Pedersoli}, \citenamefont {Savchenko}, \citenamefont {Watts}, \citenamefont
  {Zangrando}, \citenamefont {David}, \citenamefont {Capotondi}, \citenamefont
  {Vodungbo},\ and\ \citenamefont {Jal}}]{hennes_time-resolved_2021}%
  \BibitemOpen
  \bibfield  {author} {\bibinfo {author} {\bibfnamefont {M.}~\bibnamefont
  {Hennes}}, \bibinfo {author} {\bibfnamefont {B.}~\bibnamefont {Rösner}},
  \bibinfo {author} {\bibfnamefont {V.}~\bibnamefont {Chardonnet}}, \bibinfo
  {author} {\bibfnamefont {G.~S.}\ \bibnamefont {Chiuzbaian}}, \bibinfo
  {author} {\bibfnamefont {R.}~\bibnamefont {Delaunay}}, \bibinfo {author}
  {\bibfnamefont {F.}~\bibnamefont {Döring}}, \bibinfo {author} {\bibfnamefont
  {V.~A.}\ \bibnamefont {Guzenko}}, \bibinfo {author} {\bibfnamefont
  {M.}~\bibnamefont {Hehn}}, \bibinfo {author} {\bibfnamefont {R.}~\bibnamefont
  {Jarrier}}, \bibinfo {author} {\bibfnamefont {A.}~\bibnamefont {Kleibert}},
  \bibinfo {author} {\bibfnamefont {M.}~\bibnamefont {Lebugle}}, \bibinfo
  {author} {\bibfnamefont {J.}~\bibnamefont {Lüning}}, \bibinfo {author}
  {\bibfnamefont {G.}~\bibnamefont {Malinowski}}, \bibinfo {author}
  {\bibfnamefont {A.}~\bibnamefont {Merhe}}, \bibinfo {author} {\bibfnamefont
  {D.}~\bibnamefont {Naumenko}}, \bibinfo {author} {\bibfnamefont {I.~P.}\
  \bibnamefont {Nikolov}}, \bibinfo {author} {\bibfnamefont {I.}~\bibnamefont
  {Lopez-Quintas}}, \bibinfo {author} {\bibfnamefont {E.}~\bibnamefont
  {Pedersoli}}, \bibinfo {author} {\bibfnamefont {T.}~\bibnamefont
  {Savchenko}}, \bibinfo {author} {\bibfnamefont {B.}~\bibnamefont {Watts}},
  \bibinfo {author} {\bibfnamefont {M.}~\bibnamefont {Zangrando}}, \bibinfo
  {author} {\bibfnamefont {C.}~\bibnamefont {David}}, \bibinfo {author}
  {\bibfnamefont {F.}~\bibnamefont {Capotondi}}, \bibinfo {author}
  {\bibfnamefont {B.}~\bibnamefont {Vodungbo}}, \ and\ \bibinfo {author}
  {\bibfnamefont {E.}~\bibnamefont {Jal}},\ }\href {\doibase
  10.3390/app11010325} {\bibfield  {journal} {\bibinfo  {journal} {Applied
  Sciences}\ }\textbf {\bibinfo {volume} {11}},\ \bibinfo {pages} {325}
  (\bibinfo {year} {2021})}\BibitemShut {NoStop}%
\bibitem [{\citenamefont {Probst}\ \emph {et~al.}(2023)\citenamefont {Probst},
  \citenamefont {Möller}, \citenamefont {Schumacher}, \citenamefont {Brede},
  \citenamefont {Dewhurst}, \citenamefont {Reutzel}, \citenamefont {Steil},
  \citenamefont {Sharma}, \citenamefont {Jansen},\ and\ \citenamefont
  {Mathias}}]{probst_unraveling_2023}%
  \BibitemOpen
  \bibfield  {author} {\bibinfo {author} {\bibfnamefont {H.}~\bibnamefont
  {Probst}}, \bibinfo {author} {\bibfnamefont {C.}~\bibnamefont {Möller}},
  \bibinfo {author} {\bibfnamefont {M.}~\bibnamefont {Schumacher}}, \bibinfo
  {author} {\bibfnamefont {T.}~\bibnamefont {Brede}}, \bibinfo {author}
  {\bibfnamefont {J.~K.}\ \bibnamefont {Dewhurst}}, \bibinfo {author}
  {\bibfnamefont {M.}~\bibnamefont {Reutzel}}, \bibinfo {author} {\bibfnamefont
  {D.}~\bibnamefont {Steil}}, \bibinfo {author} {\bibfnamefont
  {S.}~\bibnamefont {Sharma}}, \bibinfo {author} {\bibfnamefont {G.~S.~M.}\
  \bibnamefont {Jansen}}, \ and\ \bibinfo {author} {\bibfnamefont
  {S.}~\bibnamefont {Mathias}},\ }\href {\doibase 10.48550/arXiv.2306.02783}
  {\enquote {\bibinfo {title} {Unraveling {Femtosecond} {Spin} and {Charge}
  {Dynamics} with {EUV} {T}-{MOKE} {Spectroscopy}},}\ } (\bibinfo {year}
  {2023}),\ \bibinfo {note} {arXiv:2306.02783 [cond-mat]}\BibitemShut {NoStop}%
\bibitem [{\citenamefont {Mathias}\ \emph {et~al.}(2012)\citenamefont
  {Mathias}, \citenamefont {La-O-Vorakiat}, \citenamefont {Grychtol},
  \citenamefont {Granitzka}, \citenamefont {Turgut}, \citenamefont {Shaw},
  \citenamefont {Adam}, \citenamefont {Nembach}, \citenamefont {Siemens},
  \citenamefont {Eich}, \citenamefont {Schneider}, \citenamefont {Silva},
  \citenamefont {Aeschlimann}, \citenamefont {Murnane},\ and\ \citenamefont
  {Kapteyn}}]{Mathias2012}%
  \BibitemOpen
  \bibfield  {author} {\bibinfo {author} {\bibfnamefont {S.}~\bibnamefont
  {Mathias}}, \bibinfo {author} {\bibfnamefont {C.}~\bibnamefont
  {La-O-Vorakiat}}, \bibinfo {author} {\bibfnamefont {P.}~\bibnamefont
  {Grychtol}}, \bibinfo {author} {\bibfnamefont {P.}~\bibnamefont {Granitzka}},
  \bibinfo {author} {\bibfnamefont {E.}~\bibnamefont {Turgut}}, \bibinfo
  {author} {\bibfnamefont {J.~M.}\ \bibnamefont {Shaw}}, \bibinfo {author}
  {\bibfnamefont {R.}~\bibnamefont {Adam}}, \bibinfo {author} {\bibfnamefont
  {H.~T.}\ \bibnamefont {Nembach}}, \bibinfo {author} {\bibfnamefont {M.~E.}\
  \bibnamefont {Siemens}}, \bibinfo {author} {\bibfnamefont {S.}~\bibnamefont
  {Eich}}, \bibinfo {author} {\bibfnamefont {C.~M.}\ \bibnamefont {Schneider}},
  \bibinfo {author} {\bibfnamefont {T.~J.}\ \bibnamefont {Silva}}, \bibinfo
  {author} {\bibfnamefont {M.}~\bibnamefont {Aeschlimann}}, \bibinfo {author}
  {\bibfnamefont {M.~M.}\ \bibnamefont {Murnane}}, \ and\ \bibinfo {author}
  {\bibfnamefont {H.~C.}\ \bibnamefont {Kapteyn}},\ }\href {\doibase
  10.1073/pnas.1201371109} {\bibfield  {journal} {\bibinfo  {journal}
  {Proceedings of the National Academy of Sciences}\ }\textbf {\bibinfo
  {volume} {109}},\ \bibinfo {pages} {4792} (\bibinfo {year}
  {2012})}\BibitemShut {NoStop}%
\bibitem [{\citenamefont {G\"unther}\ \emph {et~al.}(2014)\citenamefont
  {G\"unther}, \citenamefont {Spezzani}, \citenamefont {Ciprian}, \citenamefont
  {Grazioli}, \citenamefont {Ressel}, \citenamefont {Coreno}, \citenamefont
  {Poletto}, \citenamefont {Miotti}, \citenamefont {Sacchi}, \citenamefont
  {Panaccione}, \citenamefont {Uhl\'{\i}\ifmmode~\check{r}\else \v{r}\fi{}},
  \citenamefont {Fullerton}, \citenamefont {De~Ninno},\ and\ \citenamefont
  {Back}}]{Guenther2014}%
  \BibitemOpen
  \bibfield  {author} {\bibinfo {author} {\bibfnamefont {S.}~\bibnamefont
  {G\"unther}}, \bibinfo {author} {\bibfnamefont {C.}~\bibnamefont {Spezzani}},
  \bibinfo {author} {\bibfnamefont {R.}~\bibnamefont {Ciprian}}, \bibinfo
  {author} {\bibfnamefont {C.}~\bibnamefont {Grazioli}}, \bibinfo {author}
  {\bibfnamefont {B.}~\bibnamefont {Ressel}}, \bibinfo {author} {\bibfnamefont
  {M.}~\bibnamefont {Coreno}}, \bibinfo {author} {\bibfnamefont
  {L.}~\bibnamefont {Poletto}}, \bibinfo {author} {\bibfnamefont
  {P.}~\bibnamefont {Miotti}}, \bibinfo {author} {\bibfnamefont
  {M.}~\bibnamefont {Sacchi}}, \bibinfo {author} {\bibfnamefont
  {G.}~\bibnamefont {Panaccione}}, \bibinfo {author} {\bibfnamefont {V.~c.~v.}\
  \bibnamefont {Uhl\'{\i}\ifmmode~\check{r}\else \v{r}\fi{}}}, \bibinfo
  {author} {\bibfnamefont {E.~E.}\ \bibnamefont {Fullerton}}, \bibinfo {author}
  {\bibfnamefont {G.}~\bibnamefont {De~Ninno}}, \ and\ \bibinfo {author}
  {\bibfnamefont {C.~H.}\ \bibnamefont {Back}},\ }\href {\doibase
  10.1103/PhysRevB.90.180407} {\bibfield  {journal} {\bibinfo  {journal} {Phys.
  Rev. B}\ }\textbf {\bibinfo {volume} {90}},\ \bibinfo {pages} {180407}
  (\bibinfo {year} {2014})}\BibitemShut {NoStop}%
\bibitem [{\citenamefont {Yao}\ \emph {et~al.}(2020)\citenamefont {Yao},
  \citenamefont {Willems}, \citenamefont {von Korff~Schmising}, \citenamefont
  {Radu}, \citenamefont {Strüber}, \citenamefont {Schick}, \citenamefont
  {Engel}, \citenamefont {Tsukamoto}, \citenamefont {Dewhurst}, \citenamefont
  {Sharma},\ and\ \citenamefont {{others}}}]{yao_distinct_2020}%
  \BibitemOpen
  \bibfield  {author} {\bibinfo {author} {\bibfnamefont {K.}~\bibnamefont
  {Yao}}, \bibinfo {author} {\bibfnamefont {F.}~\bibnamefont {Willems}},
  \bibinfo {author} {\bibfnamefont {C.}~\bibnamefont {von Korff~Schmising}},
  \bibinfo {author} {\bibfnamefont {I.}~\bibnamefont {Radu}}, \bibinfo {author}
  {\bibfnamefont {C.}~\bibnamefont {Strüber}}, \bibinfo {author}
  {\bibfnamefont {D.}~\bibnamefont {Schick}}, \bibinfo {author} {\bibfnamefont
  {D.}~\bibnamefont {Engel}}, \bibinfo {author} {\bibfnamefont
  {A.}~\bibnamefont {Tsukamoto}}, \bibinfo {author} {\bibfnamefont
  {J.}~\bibnamefont {Dewhurst}}, \bibinfo {author} {\bibfnamefont
  {S.}~\bibnamefont {Sharma}}, \ and\ \bibinfo {author} {\bibnamefont
  {{others}}},\ }\href@noop {} {\bibfield  {journal} {\bibinfo  {journal}
  {Physical Review B}\ }\textbf {\bibinfo {volume} {102}},\ \bibinfo {pages}
  {100405} (\bibinfo {year} {2020})}\BibitemShut {NoStop}%
\bibitem [{\citenamefont {Jana}\ \emph {et~al.}(2017)\citenamefont {Jana},
  \citenamefont {Terschlüsen}, \citenamefont {Stefanuik}, \citenamefont
  {Plogmaker}, \citenamefont {Troisi}, \citenamefont {Malik}, \citenamefont
  {Svanqvist}, \citenamefont {Knut}, \citenamefont {Söderström},\ and\
  \citenamefont {Karis}}]{Jana2017}%
  \BibitemOpen
  \bibfield  {author} {\bibinfo {author} {\bibfnamefont {S.}~\bibnamefont
  {Jana}}, \bibinfo {author} {\bibfnamefont {J.~A.}\ \bibnamefont
  {Terschlüsen}}, \bibinfo {author} {\bibfnamefont {R.}~\bibnamefont
  {Stefanuik}}, \bibinfo {author} {\bibfnamefont {S.}~\bibnamefont
  {Plogmaker}}, \bibinfo {author} {\bibfnamefont {S.}~\bibnamefont {Troisi}},
  \bibinfo {author} {\bibfnamefont {R.~S.}\ \bibnamefont {Malik}}, \bibinfo
  {author} {\bibfnamefont {M.}~\bibnamefont {Svanqvist}}, \bibinfo {author}
  {\bibfnamefont {R.}~\bibnamefont {Knut}}, \bibinfo {author} {\bibfnamefont
  {J.}~\bibnamefont {Söderström}}, \ and\ \bibinfo {author} {\bibfnamefont
  {O.}~\bibnamefont {Karis}},\ }\href {\doibase 10.1063/1.4978907} {\bibfield
  {journal} {\bibinfo  {journal} {Review of Scientific Instruments}\ }\textbf
  {\bibinfo {volume} {88}},\ \bibinfo {pages} {033113} (\bibinfo {year}
  {2017})}\BibitemShut {NoStop}%
\bibitem [{\citenamefont {Möller}\ \emph {et~al.}(2021)\citenamefont
  {Möller}, \citenamefont {Probst}, \citenamefont {Otto}, \citenamefont
  {Stroh}, \citenamefont {Mahn}, \citenamefont {Steil}, \citenamefont
  {Moshnyaga}, \citenamefont {Jansen}, \citenamefont {Steil},\ and\
  \citenamefont {Mathias}}]{moller_ultrafast_2021}%
  \BibitemOpen
  \bibfield  {author} {\bibinfo {author} {\bibfnamefont {C.}~\bibnamefont
  {Möller}}, \bibinfo {author} {\bibfnamefont {H.}~\bibnamefont {Probst}},
  \bibinfo {author} {\bibfnamefont {J.}~\bibnamefont {Otto}}, \bibinfo {author}
  {\bibfnamefont {K.}~\bibnamefont {Stroh}}, \bibinfo {author} {\bibfnamefont
  {C.}~\bibnamefont {Mahn}}, \bibinfo {author} {\bibfnamefont {S.}~\bibnamefont
  {Steil}}, \bibinfo {author} {\bibfnamefont {V.}~\bibnamefont {Moshnyaga}},
  \bibinfo {author} {\bibfnamefont {G.~M.}\ \bibnamefont {Jansen}}, \bibinfo
  {author} {\bibfnamefont {D.}~\bibnamefont {Steil}}, \ and\ \bibinfo {author}
  {\bibfnamefont {S.}~\bibnamefont {Mathias}},\ }\href@noop {} {\bibfield
  {journal} {\bibinfo  {journal} {Review of Scientific Instruments}\ }\textbf
  {\bibinfo {volume} {92}},\ \bibinfo {pages} {065107} (\bibinfo {year}
  {2021})}\BibitemShut {NoStop}%
\bibitem [{\citenamefont {Jana}\ \emph {et~al.}(2022)\citenamefont {Jana},
  \citenamefont {Knut}, \citenamefont {Muralidhar}, \citenamefont {Malik},
  \citenamefont {Stefanuik}, \citenamefont {Åkerman}, \citenamefont {Karis},
  \citenamefont {Schüßler-Langeheine},\ and\ \citenamefont
  {Pontius}}]{Jana2022}%
  \BibitemOpen
  \bibfield  {author} {\bibinfo {author} {\bibfnamefont {S.}~\bibnamefont
  {Jana}}, \bibinfo {author} {\bibfnamefont {R.}~\bibnamefont {Knut}}, \bibinfo
  {author} {\bibfnamefont {S.}~\bibnamefont {Muralidhar}}, \bibinfo {author}
  {\bibfnamefont {R.~S.}\ \bibnamefont {Malik}}, \bibinfo {author}
  {\bibfnamefont {R.}~\bibnamefont {Stefanuik}}, \bibinfo {author}
  {\bibfnamefont {J.}~\bibnamefont {Åkerman}}, \bibinfo {author}
  {\bibfnamefont {O.}~\bibnamefont {Karis}}, \bibinfo {author} {\bibfnamefont
  {C.}~\bibnamefont {Schüßler-Langeheine}}, \ and\ \bibinfo {author}
  {\bibfnamefont {N.}~\bibnamefont {Pontius}},\ }\href {\doibase
  10.1063/5.0080331} {\bibfield  {journal} {\bibinfo  {journal} {Applied
  Physics Letters}\ }\textbf {\bibinfo {volume} {120}},\ \bibinfo {pages}
  {102404} (\bibinfo {year} {2022})}\BibitemShut {NoStop}%
\bibitem [{\citenamefont {Dewhurst}\ \emph {et~al.}(2016)\citenamefont
  {Dewhurst}, \citenamefont {Krieger}, \citenamefont {Sharma},\ and\
  \citenamefont {Gross}}]{Dewhurst.2016}%
  \BibitemOpen
  \bibfield  {author} {\bibinfo {author} {\bibfnamefont {J.}~\bibnamefont
  {Dewhurst}}, \bibinfo {author} {\bibfnamefont {K.}~\bibnamefont {Krieger}},
  \bibinfo {author} {\bibfnamefont {S.}~\bibnamefont {Sharma}}, \ and\ \bibinfo
  {author} {\bibfnamefont {E.}~\bibnamefont {Gross}},\ }\href {\doibase
  10.1016/j.cpc.2016.09.001} {\bibfield  {journal} {\bibinfo  {journal}
  {Computer Physics Communications}\ }\textbf {\bibinfo {volume} {209}},\
  \bibinfo {pages} {92} (\bibinfo {year} {2016})}\BibitemShut {NoStop}%
\bibitem [{\citenamefont {Krieger}\ \emph {et~al.}(2015)\citenamefont
  {Krieger}, \citenamefont {Dewhurst}, \citenamefont {Elliott}, \citenamefont
  {Sharma},\ and\ \citenamefont {Gross}}]{krieger2015laser}%
  \BibitemOpen
  \bibfield  {author} {\bibinfo {author} {\bibfnamefont {K.}~\bibnamefont
  {Krieger}}, \bibinfo {author} {\bibfnamefont {J.}~\bibnamefont {Dewhurst}},
  \bibinfo {author} {\bibfnamefont {P.}~\bibnamefont {Elliott}}, \bibinfo
  {author} {\bibfnamefont {S.}~\bibnamefont {Sharma}}, \ and\ \bibinfo {author}
  {\bibfnamefont {E.}~\bibnamefont {Gross}},\ }\href@noop {} {\bibfield
  {journal} {\bibinfo  {journal} {Journal of chemical theory and computation}\
  }\textbf {\bibinfo {volume} {11}},\ \bibinfo {pages} {4870} (\bibinfo {year}
  {2015})}\BibitemShut {NoStop}%
\bibitem [{\citenamefont {Bierbrauer}\ \emph {et~al.}(2017)\citenamefont
  {Bierbrauer}, \citenamefont {Weber}, \citenamefont {Schummer}, \citenamefont
  {Barkowski}, \citenamefont {Mahro}, \citenamefont {Mathias}, \citenamefont
  {Schneider}, \citenamefont {Stadtmüller}, \citenamefont {Aeschlimann},\ and\
  \citenamefont {Rethfeld}}]{Bierbrauer:2017db}%
  \BibitemOpen
  \bibfield  {author} {\bibinfo {author} {\bibfnamefont {U.}~\bibnamefont
  {Bierbrauer}}, \bibinfo {author} {\bibfnamefont {S.~T.}\ \bibnamefont
  {Weber}}, \bibinfo {author} {\bibfnamefont {D.}~\bibnamefont {Schummer}},
  \bibinfo {author} {\bibfnamefont {M.}~\bibnamefont {Barkowski}}, \bibinfo
  {author} {\bibfnamefont {A.-K.}\ \bibnamefont {Mahro}}, \bibinfo {author}
  {\bibfnamefont {S.}~\bibnamefont {Mathias}}, \bibinfo {author} {\bibfnamefont
  {H.~C.}\ \bibnamefont {Schneider}}, \bibinfo {author} {\bibfnamefont
  {B.}~\bibnamefont {Stadtmüller}}, \bibinfo {author} {\bibfnamefont
  {M.}~\bibnamefont {Aeschlimann}}, \ and\ \bibinfo {author} {\bibfnamefont
  {B.}~\bibnamefont {Rethfeld}},\ }\href@noop {} {\bibfield  {journal}
  {\bibinfo  {journal} {Journal of Physics: Condensed Matter}\ }\textbf
  {\bibinfo {volume} {29}},\ \bibinfo {pages} {244002} (\bibinfo {year}
  {2017})}\BibitemShut {NoStop}%
\bibitem [{\citenamefont {Dewhurst}\ \emph {et~al.}(2020)\citenamefont
  {Dewhurst}, \citenamefont {Willems}, \citenamefont {Elliott}, \citenamefont
  {Li}, \citenamefont {Schmising}, \citenamefont {Strüber}, \citenamefont
  {Engel}, \citenamefont {Eisebitt},\ and\ \citenamefont
  {Sharma}}]{Dewhurst2020PRL}%
  \BibitemOpen
  \bibfield  {author} {\bibinfo {author} {\bibfnamefont {J.~K.}\ \bibnamefont
  {Dewhurst}}, \bibinfo {author} {\bibfnamefont {F.}~\bibnamefont {Willems}},
  \bibinfo {author} {\bibfnamefont {P.}~\bibnamefont {Elliott}}, \bibinfo
  {author} {\bibfnamefont {Q.~Z.}\ \bibnamefont {Li}}, \bibinfo {author}
  {\bibfnamefont {C.~v.~K.}\ \bibnamefont {Schmising}}, \bibinfo {author}
  {\bibfnamefont {C.}~\bibnamefont {Strüber}}, \bibinfo {author}
  {\bibfnamefont {D.~W.}\ \bibnamefont {Engel}}, \bibinfo {author}
  {\bibfnamefont {S.}~\bibnamefont {Eisebitt}}, \ and\ \bibinfo {author}
  {\bibfnamefont {S.}~\bibnamefont {Sharma}},\ }\href {\doibase
  10.1103/PhysRevLett.124.077203} {\bibfield  {journal} {\bibinfo  {journal}
  {Physical Review Letters}\ }\textbf {\bibinfo {volume} {124}},\ \bibinfo
  {pages} {077203} (\bibinfo {year} {2020})},\ \Eprint
  {http://arxiv.org/abs/1909.00199} {1909.00199} \BibitemShut {NoStop}%
\bibitem [{\citenamefont {Dewhurst}\ \emph {et~al.}()\citenamefont {Dewhurst},
  \citenamefont {Sharma},\ and\ \citenamefont {et. al.}}]{dewhurstsharma}%
  \BibitemOpen
  \bibfield  {author} {\bibinfo {author} {\bibfnamefont {J.~K.}\ \bibnamefont
  {Dewhurst}}, \bibinfo {author} {\bibfnamefont {S.}~\bibnamefont {Sharma}}, \
  and\ \bibinfo {author} {\bibnamefont {et. al.}},\ }\href@noop {} {}\bibinfo
  {howpublished} {\url{elk.sourceforge.net}}\BibitemShut {NoStop}%
\bibitem [{\citenamefont {Koopmans}\ \emph {et~al.}(2010)\citenamefont
  {Koopmans}, \citenamefont {Malinowski}, \citenamefont {Longa}, \citenamefont
  {Steiauf}, \citenamefont {Fähnle}, \citenamefont {Roth}, \citenamefont
  {Cinchetti},\ and\ \citenamefont {Aeschlimann}}]{Koopmans:2010eu}%
  \BibitemOpen
  \bibfield  {author} {\bibinfo {author} {\bibfnamefont {B.}~\bibnamefont
  {Koopmans}}, \bibinfo {author} {\bibfnamefont {G.}~\bibnamefont
  {Malinowski}}, \bibinfo {author} {\bibfnamefont {F.~D.}\ \bibnamefont
  {Longa}}, \bibinfo {author} {\bibfnamefont {D.}~\bibnamefont {Steiauf}},
  \bibinfo {author} {\bibfnamefont {M.}~\bibnamefont {Fähnle}}, \bibinfo
  {author} {\bibfnamefont {T.}~\bibnamefont {Roth}}, \bibinfo {author}
  {\bibfnamefont {M.}~\bibnamefont {Cinchetti}}, \ and\ \bibinfo {author}
  {\bibfnamefont {M.}~\bibnamefont {Aeschlimann}},\ }\href {\doibase
  10.1038/nmat2593} {\bibfield  {journal} {\bibinfo  {journal} {Nature
  Materials}\ }\textbf {\bibinfo {volume} {9}},\ \bibinfo {pages} {259}
  (\bibinfo {year} {2010})}\BibitemShut {NoStop}%
\bibitem [{Note1()}]{Note1}%
  \BibitemOpen
  \bibinfo {note} {This data is reproduced from Ref.~\cite
  {probst_unraveling_2023}, where we describe the \protect \ensuremath
  {\epsilon _{xy}}{} analysis procedure in detail}\BibitemShut {NoStop}%
\end{thebibliography}%
